\documentclass[12pt]{article}
\usepackage[margin=1in]{geometry}
\usepackage[numbers,square,sort&compress]{natbib}%
\usepackage{amsmath,amssymb}
\usepackage[utf8]{inputenc}
\usepackage[T1]{fontenc}
\usepackage{lmodern}
\usepackage[hyphens]{url}
\usepackage{hyperref}\hypersetup{hidelinks}
\usepackage{graphicx}
\usepackage{booktabs}
\usepackage{longtable}
\usepackage{array}
\usepackage{caption}
\usepackage[most]{tcolorbox}          %
\newtcolorbox[auto counter]{biobox}[1][]{%
  breakable, lines before break=4, colback=black!3, colframe=black!45, boxrule=0.4pt, arc=1pt,
  fonttitle=\bfseries, coltitle=black, left=4pt, right=4pt, top=3pt, bottom=3pt,
  before upper={\setlength{\parskip}{0.6ex plus 0.2ex}},%
  title={Box~\thetcbcounter\ --- theoretical details}, #1}
\tcbset{coltitle=black}%
\newcolumntype{L}[1]{>{\raggedright\arraybackslash}p{#1}}%
\newcommand{\indep}{\mathrel{\perp\!\!\!\perp}}%
\usepackage{setspace}
\usepackage{etoolbox}
\patchcmd{\abstract}{\small}{\normalsize}{}{\errmessage{the abstract font patch failed}}%

\providecommand{\tightlist}{\setlength{\itemsep}{0pt}\setlength{\parskip}{0pt}}

\newenvironment{keywords}{\par\medskip\noindent\textbf{Keywords:} }{\par\medskip}

\title{Three routes to one answer: reconciling AIPW, TMLE, and double machine learning for~applied~researchers}

\author{M. Ehsan Karim\\[4pt]
\normalsize
\begin{minipage}{0.9\textwidth}\centering
School of Population and Public Health, University of British Columbia, Vancouver,
British Columbia, Canada;\\ and Centre for Advancing Health Outcomes, St.\ Paul's
Hospital, Vancouver, British Columbia, Canada\\[3pt]
\texttt{ehsan.karim@ubc.ca}\\[2pt]
ORCID: 0000-0002-0346-2871
\end{minipage}}
\date{}

\begin{document}
\nocite{naimi2023,mooney2021,chernozhukov2018,diaz2020,qiu2024,robins1994,bickel1993,zhong2021,vanderlaan2006,vanderlaan2007,luque2018,schuler2017,bach2024,kennedy2023,hines2022,gruber2010,gruber2012,coyle2021tmle3,tan2022,ellul2025,meng2026refine2,karimfu2026,moccia2024,mondol2025,hill2011,lalonde1986,dehejia1999,petersen2012,renson2025,hernan2020,zheng2011,newey2018,zivich2021,polley2024superlearner,phillips2023,karim2025splits,albert1984,connors1996,balzer2016,zivich2024seed,schader2024seed,naimi2024seed,smith2022,talbot2025,keele2025,rytgaard2023,benkeser_survtmle,chen_concrete,morris2019}
\maketitle

\begin{abstract}
Augmented inverse probability weighting (AIPW), targeted maximum likelihood estimation
(TMLE), and double/debiased machine learning (DML) are three asymptotically equivalent
routes to the same efficient influence function for the average treatment effect. We
contribute a worked, shared-nuisance reconciliation on real data: what a practitioner must
hold fixed for the routes and their software to agree, and when they do not. Using open data
on smoking cessation and weight change (\textit{n}~=~1566), one Super Learner library, and
folds matched within schemes, we build all three by hand, in full-sample, cross-fit, and
double-cross-fit variants. A semi-synthetic benchmark with known sample-average effects on a
fixed design grades interval coverage. The six doubly robust estimates span 3.32--3.39~kg
over repeated splits. Single runs of our code and the \texttt{tmle}, \texttt{AIPW},
\texttt{DoubleML}, and \texttt{tmle3} packages span 3.32--3.49~kg, compatible with
fitting-scheme differences and single-split noise. Over 20 seeds, the three implementations
sharing our library agree within 0.01~kg in median, although one is a TMLE. Agreement can
fail: with in-sample nuisance fits on a binary mortality outcome, and under poor overlap in
a job-training example. We recommend diagnosing overlap before choosing an estimator, and
provide a reporting checklist and software code.
\end{abstract}

\par\medskip

\setstretch{1.05}
\section{Introduction}\label{introduction}

\subsection{Why an applied researcher reaches for a doubly-robust
estimator}\label{why-an-applied-researcher-reaches-for-a-doubly-robust-estimator}

Analysts increasingly use flexible machine-learning (ML) algorithms to model treatment assignment and outcomes using covariates selected for confounding adjustment \citep{naimi2023, mooney2021}.
Examples include random forests, gradient boosting, penalized regression, and ensembles.
The appeal is plain: we no longer have to guess the functional form of a propensity score or an outcome regression.
The hazard is less plain.
Plugging ML fits directly into g-computation or inverse-probability weighting can leave first-order nuisance-estimation bias \citep{chernozhukov2018, diaz2020, qiu2024}.
An interval that ignores this bias can undercover (Box~\ref{box:b1}).
The shaded, numbered theory boxes carry the supporting semiparametric detail and can be skipped on a first reading; the main text stands alone.

\begin{biobox}[label={box:b1}, unbreakable]
This box shows why a flexible outcome-regression plug-in can carry first-order bias, and how the doubly-robust estimators replace it.
Let $O=(X,A,Y)\sim P$, let $\psi=E[Y^{1}-Y^{0}]$ be the average treatment effect (Step~0), and let $g(X)=P(A=1\mid X)$ be the propensity score.
Let $\hat Q_a$ be a data-adaptive estimate of $Q_a(X)=E[Y\mid A=a,X]$, and let $\mathbb{P}_n$ be the empirical mean.
The outcome-regression plug-in (g-computation) estimator is $\hat\psi_{\mathrm{plug}} = \mathbb{P}_n\{\hat Q_1(X)-\hat Q_0(X)\}$. Its error decomposes as
$$
\begin{aligned}
\hat\psi_{\mathrm{plug}}-\psi
={}& \underbrace{(\mathbb{P}_n-P)\{Q_1-Q_0\}}_{O_P(n^{-1/2})}
\;+\; \underbrace{P\{(\hat Q_1-Q_1)-(\hat Q_0-Q_0)\}}_{\text{first-order (plug-in) bias}}
\\ &\quad+\; \underbrace{(\mathbb{P}_n-P)\{(\hat Q_1-Q_1)-(\hat Q_0-Q_0)\}}_{\text{empirical-process term}}.
\end{aligned}
$$
The last term is $o_P(n^{-1/2})$ when $\hat Q$ is consistent in $L_2(P)$ and a Donsker condition or the cross-fitting of Steps~\mbox{3--5} controls it.
The middle term is the first-order plug-in bias.
Its absolute value is bounded by $\|\hat Q_1-Q_1\|+\|\hat Q_0-Q_0\|$, but it need not be that large, because errors can cancel when averaged.

Flexible ML estimators typically converge slower than $n^{-1/2}$, so nothing makes this bias negligible at the $n^{-1/2}$ scale.
Standard root-$n$ inference for the plug-in is therefore not justified, and its nominal $95\%$ interval can undercover.

The doubly-robust estimators of Steps~3--4 replace this first-order term with a second-order remainder bounded by the product of nuisance errors $\|\hat g-g\|\,\|\hat Q-Q\|$ (Box~\ref{box:b6}).
This remainder is $o_P(n^{-1/2})$ whenever the two nuisance rates multiply to faster than $n^{-1/2}$, for instance when each is $o_P(n^{-1/4})$.
This vanishing of first-order nuisance dependence is the Neyman-orthogonality property of the full efficient-influence-function moment.
AIPW and DML use that moment, and TMLE targets it \citep{robins1994,bickel1993,chernozhukov2018}.
\end{biobox}

\subsection{Three literatures, one estimating equation}\label{three-literatures-one-estimator}

An analyst who goes looking for the fix meets it three times, under
three vocabularies, in three software ecosystems:

\begin{itemize}
\tightlist
\item
  \textbf{Augmented inverse-probability weighting (AIPW / AIPTW)} ---
  the one-step estimator from the missing-data and the
  semiparametric-efficiency traditions \citep{robins1994, zhong2021}.
\item
  \textbf{Targeted maximum likelihood estimation (TMLE)} --- the plug-in
  that \emph{targets} the parameter, from the targeted-learning
  tradition
  \citep{vanderlaan2006, vanderlaan2007, luque2018, schuler2017}.
\item
  \textbf{Double / debiased machine learning (DML)} --- the
  orthogonal-moment-with-cross-fitting estimator from the econometrics
  tradition \citep{chernozhukov2018, bach2024}.
\end{itemize}

On the surface they look unrelated: an ``augmentation term,'' a ``clever
covariate and fluctuation,'' a ``Neyman-orthogonal score.'' They ship in
different packages with different defaults. A practitioner is in a
genuinely awkward spot when they run one package for TMLE and another
for DML on the same data, read two different numbers, and cannot tell
whether the gap is \emph{the estimator} or \emph{the implementation}.

The settled theory says that, in large samples, the gap is not the estimator. All three are routes to the same efficient influence function (EIF) for the average treatment effect (ATE). They solve the same estimating equation. Under the conditions of Box~\ref{box:b11}, they are asymptotically equivalent and efficient \citep{kennedy2023, hines2022, diaz2020}. We treat that equivalence as
established background, not as a result of this tutorial. What confuses
practitioners is everything that sits \emph{on top of} the theory: the
words, the code defaults, and the finite-sample choices (cross-fitting,
fold count, propensity truncation) that make two correct estimators
print two different numbers.

\subsection{What this tutorial does}\label{what-this-tutorial-does}

This tutorial connects a common ATE score to the fitting and combination choices used in practice.
We work through an open NHEFS analysis with shared nuisance fits, compare five R implementations, and develop a reporting checklist.
The comparisons work at two levels: matched comparisons within our engine, and descriptive comparisons across packages.
Concretely:

\begin{itemize}
\tightlist
\item
  \textbf{Common foundation.} AIPW, TMLE, and DML are presented as three
  routes to the same EIF for the ATE, rather than three unrelated
  recipes (Steps~3--4).
\item
  \textbf{Matched within-engine comparisons.} Every estimator is computed on the same open NHEFS smoking-cessation data, with a shared nuisance library. Within each fitting scheme the estimators also share folds and initial nuisance fits, so a difference between them isolates the combination rule rather than reflecting how two packages were configured.
\item
    \textbf{A software bridge.} We compare the hand-coded estimators with the major R implementations on the identical estimand: \texttt{tmle} \citep{gruber2010, gruber2012}, \texttt{AIPW} \citep{zhong2021}, \texttt{DoubleML} \citep{bach2024}, and \texttt{tmle3} \citep{coyle2021tmle3}.
  Several settings vary together across packages, so these comparisons are descriptive: they quantify agreement and variability without identifying each setting's separate contribution.
  Run once each, the five implementations (ours and the four packages) differ by up to 0.18 kg.
  That is about as much as our own six doubly-robust estimators differ on a single split (median span 0.15 kg; 0.07 kg with repeated splits).
  Three implementations share our \texttt{SuperLearner} library: the hand-coded engine, \texttt{tmle}, and \texttt{AIPW}.
  For these three, we find that spread compatible with the fitting scheme (in-sample or out-of-fold) plus single-split noise, without appeal to the estimator label (Step~7).
\item
  \textbf{A reporting checklist.} The walkthrough is condensed into a
  short list of what to pre-specify and report.
\end{itemize}

None of these is a new estimator or a new theorem.
Simulation studies have already compared how AIPW and TMLE perform with machine-learning nuisances \citep{tan2022, ellul2025, meng2026refine2, karimfu2026}.
What this tutorial contributes instead is the reconciliation itself, worked through with shared nuisance fits and graded against a benchmark.
It complements the code-free review of \citet{moccia2024} and the single-estimator guide of \citet{mondol2025} by working all three routes through one code path (Discussion).
The software bridge is the piece most directly useful in applied practice.
Three further examples have distinct jobs.
Right-heart catheterization (RHC) data, with a binary outcome, show that one-step and targeted estimates can differ when the initial fits are in sample (Step~4).
They also support a study of repeated splitting (Step~5).
A semi-synthetic benchmark \citep{hill2011} checks the routes against known sample-average effects on a fixed design (Step~5).
A positivity-failure vignette on the LaLonde/NSW--PSID data \citep{lalonde1986, dehejia1999} shows how they behave when overlap fails (Step~6).

The organizing thesis is a single instruction that we place ahead of the
choice of estimator: \textbf{diagnose overlap first}. In our examples the three routes agree closely when the treatment groups overlap on the confounders and the nuisances are fit out of fold (Step~4). Where overlap is poor, estimation becomes unstable, and in the example of Step~6 the routes diverge sharply from one another. The decision that comes first is therefore the positivity diagnosis, not the label AIPW-versus-TMLE-versus-DML
\citep{petersen2012, renson2025}. This is not a horse race, and we do
not crown a winner --- asymptotically there is none to crown.

\textbf{Learning objectives.} By the end of this tutorial you will be
able to:

\begin{enumerate}
\def\labelenumi{\arabic{enumi}.}
\tightlist
\item
  state the ATE estimand and the three assumptions that identify it, and
  diagnose overlap \emph{before} estimating;
\item
  recognise AIPW/AIPTW, TMLE, and DML as three routes to one
  efficient influence function;
\item
  fit one shared Super Learner nuisance stack and read off the
  doubly-robust estimate and its analytic standard error;
\item
  explain what cross-fitting buys, and why a coverage number needs a
  replicate count and a Monte Carlo standard error;
\item
  trace numerically different answers from \texttt{tmle}, \texttt{AIPW},
  \texttt{DoubleML}, and \texttt{tmle3} to the implementation settings behind them (fitting scheme, number of repeated splits, learner library, truncation) rather than assuming they reflect the estimator label;
  and
\item
  report a doubly-robust analysis so a reader can audit and reproduce
  it.
\end{enumerate}

\section{Step~0 --- The question, data, and
estimand}\label{step-0-the-question-data-and-estimand}

\subsection{The motivating question}\label{the-motivating-question}

Does quitting smoking change body weight? The concern is familiar in
clinical counselling: people who stop smoking often gain weight, and the
question is how much of that change is \emph{caused} by quitting rather
than by who quits. We use this question throughout as the single thread
that carries every estimator.

\subsection{The data}\label{the-data}

We use the National Health and Nutrition Examination Survey Data I
Epidemiologic Follow-up Study (NHEFS), the same open dataset used to
teach this exact question in the causal-inference literature
\citep{hernan2020}. The exposure \(A\) is quitting smoking between the 1971--75 baseline and the 1982 follow-up.
The outcome \(Y\) is the corresponding change in body weight, in kilograms.
After restricting to complete cases on the baseline confounders and the outcome, the analytic sample is \(n = 1566\), of whom \(403\) quit and \(1163\) continued smoking (Table~\ref{tab:nhefs-sample}; the companion code).
The estimand is therefore the ATE in this complete-case analytic population, and the identification assumptions of Box~\ref{box:b2} are read within it.
We do not separately model loss to follow-up.
Inclusion requires a response to the 1982 survey, after quitting began.
This can introduce selection bias.
Following \citet[Fine Point~12.1]{hernan2020}, we do not address that bias here.
The data are freely available through the \texttt{causaldata} package, so the NHEFS analyses can be rerun from public data.

This is a good-overlap example: on the estimated propensity-score scale, quitters and continuers overlap well.
Good overlap is what makes NHEFS a clean vehicle for the equivalence story.
The overlap itself is diagnosed and displayed in Step~1.

\begin{table}[ht]
\centering
\caption{The NHEFS analytic sample for the smoking-cessation question. The exposure $A$ is quitting smoking between the 1971--75 baseline and the 1982 follow-up; the outcome $Y$ is the corresponding change in body weight (kilograms).}
\label{tab:nhefs-sample}
\begin{tabular}{lr}
\toprule
Group & $n$ \\
\midrule
Quit smoking ($A=1$)      & 403 \\
Continued smoking ($A=0$) & 1163 \\
\midrule
Total analytic sample     & 1566 \\
\bottomrule
\end{tabular}
\end{table}

\subsection{The causal story: a hypothesized
DAG}\label{the-causal-story-a-hypothesized-dag}

\begin{figure}[htbp]\centering
\includegraphics[width=0.72\linewidth]{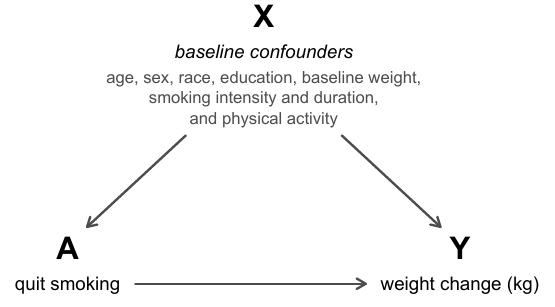}
\caption{\label{fig:dag}Hypothesized causal directed acyclic graph for the NHEFS question. Quitting smoking ($A$) affects 1971--1982 weight change ($Y$); a set of baseline covariates $X$ (age, sex, race, education, baseline weight, smoking intensity and duration, and physical activity) are common causes of both $A$ and $Y$ and form the adjustment set.}
\end{figure}

The confounding structure we are willing to assume is drawn as a
directed acyclic graph (DAG) in Figure \ref{fig:dag}. Quitting smoking
(\(A\)) is a cause of weight change (\(Y\)). A set of baseline
covariates \(X\) (age, sex, race, education, baseline weight, smoking
intensity and duration, and physical activity) are common
causes of both quitting and subsequent weight change, and constitute the
adjustment set. The DAG is a \emph{hypothesis}, not a finding: it
encodes the confounders we believe we must control and asserts there are
no others left unmeasured. Everything downstream is a way of adjusting
for exactly this \(X\).

\subsection{The estimand and what it takes to identify
it}\label{the-estimand-and-what-it-takes-to-identify-it}

The target of every estimator in this tutorial is one number, the
average treatment effect \[
\psi \;=\; E\big[Y^{1}-Y^{0}\big],
\] the mean difference in 1971--1982 weight change we would see if
\emph{everyone} quit versus if \emph{everyone} continued smoking.
Turning this contrast of unobserved potential outcomes into something
estimable from the observed data requires three assumptions, stated here
in words:

\begin{enumerate}
\def\labelenumi{\arabic{enumi}.}
\tightlist
\item
  \textbf{Consistency.} The weight change we observe for a person is the
  weight change that would occur under the smoking-cessation status they
  actually had --- there is a single, well-defined version of
  ``quitting.''
\item
  \textbf{Exchangeability (no unmeasured confounding).} Within levels of
  the measured confounders \(X\), quitters and continuers are comparable
  in their potential weight changes --- the DAG in Figure \ref{fig:dag}
  leaves no open backdoor path unadjusted.
\item
  \textbf{Positivity (overlap).} At every combination of confounders
  that occurs in the population, both quitting and continuing had a nonzero
  probability --- no one was structurally certain to quit or certain not
  to.
\end{enumerate}

The first two assumptions are about the science and the DAG; they are untestable from the data alone \citep{hernan2020}. The third is partly checkable, and
checking it is exactly the thesis of this tutorial --- \emph{diagnose
overlap first} --- which is why Step~1 is a positivity diagnosis rather
than an estimate (Box~\ref{box:b2}).

\begin{biobox}[label={box:b2}]
Let $O=(X,A,Y)\sim P$ with $A\in\{0,1\}$ and potential outcomes $Y^{0},Y^{1}$. The target is the average treatment effect
$$
\psi \;=\; E\big[Y^{1}-Y^{0}\big].
$$
Identification uses three assumptions:
\begin{align*}
&\textbf{(A1) Consistency:} && Y \;=\; Y^{A} \;=\; A\,Y^{1}+(1-A)\,Y^{0};\\*
&\textbf{(A2) Exchangeability:} && Y^{a}\indep A\mid X, \quad a\in\{0,1\};\\*
&\textbf{(A3) Positivity:} && 0 < g(X) < 1\ \text{a.s.}, \quad g(X)=P(A=1\mid X).
\end{align*}
Under (A1)--(A3), $\psi$ is identified by the g-formula
$$
\psi \;=\; E\big[\,Q_1(X)-Q_0(X)\,\big], \qquad Q_a(X)=E[Y\mid A=a,X].
$$
The identified estimand depends on $P$ through the outcome regression $Q$ and the marginal law of $X$, not through the propensity $g$.
The propensity enters the efficient-influence-function estimating equation $\mathbb{P}_n\,D(O;g,Q,\psi)=0$ that every doubly-robust estimator in this tutorial solves for this $\psi$ (Step~3 assembles $D$ explicitly). This is why g-computation needs only $Q$ while the doubly-robust routes read from both.
\end{biobox}

\subsection{Roadmap}\label{roadmap}

The rest of the tutorial is a single worked example, whose eight steps
are laid out in Table \ref{tab:steps}. Having fixed the question, data,
and estimand here (Step~0), we \textbf{diagnose overlap first} (Step~1),
fit the two nuisance functions \(g\) and \(Q\) with a Super Learner
ensemble (Step~2), and assemble the efficient influence function that
the three estimators share (Step~3). We then present all three routes to that one score: AIPW\,/\,one-step, TMLE by targeting, and DML with cross-fitting (Step~4).
Step~4 also gives double-cross-fit forms of AIPW and TMLE.
We use single cross-fit TMLE in the sense of CV-TMLE \citep{zheng2011} and double cross-fitting in the sense of \citet{newey2018} and \citet{zivich2021}.
In Step~5 we examine what cross-fitting does, and does not do, for machine-learning inference, calibrate the routes against a known-truth benchmark, and track how the estimates settle as the split is repeated on one large dataset.
In Step~6 we probe what poor overlap does and the analyst's design sensitivities.
Finally, in Step~7 we reconcile the answer across software packages while distilling a reporting checklist.

\begin{table}[htbp]\centering
\caption{The eight steps of the walkthrough. Each step is one move an analyst makes, in order, on the running NHEFS example. Later steps add RHC for a binary-outcome check, IHDP for a known-truth check, and LaLonde/NSW--PSID for a poor-overlap check.}
\label{tab:steps}
\begin{tabular}{@{}cl@{}}
\toprule
Step & What it does \\
\midrule
0 & The question, data (NHEFS), estimand, and the three identification assumptions \\
1 & Diagnose overlap first --- the positivity check that gates every estimate \\
2 & Fit both nuisances with one shared Super Learner library and matched folds \\
3 & The efficient influence function, term by term, with the singly-robust baselines \\
4 & AIPW, TMLE, and DML: two combination rules, three fitting schemes, one score \\
5 & What cross-fitting does, a known-truth calibration check, and repeating the split \\
6 & Sensitivity analyses and what poor overlap does \\
7 & A reporting checklist and reconciliation across software \\
\bottomrule
\end{tabular}
\end{table}

\textbf{Key takeaway.} Fix the estimand and its three identification
assumptions before choosing an estimator; of the three, only positivity is partly checkable from the data.

\begin{tcolorbox}[title={Step~0 in pseudocode --- question, data, estimand}, colback=blue!3, colframe=blue!35, fonttitle=\bfseries, boxrule=0.4pt, arc=1pt]
load NHEFS (\texttt{causaldata}); keep complete cases $\Rightarrow n = 1566$ \\
$A \leftarrow$ quit smoking (0/1); \quad $Y \leftarrow$ 1971--1982 weight change (kg); \\
$X \leftarrow$ baseline confounders \\
target $\psi = E[Y^{1}-Y^{0}]$ \quad \textit{\# identified under consistency, exchangeability, positivity}
\end{tcolorbox}

\section{Step~1 --- Diagnose overlap
first}\label{step-1-diagnose-overlap-first}

Of the three identification assumptions of Step~0 (Box~\ref{box:b2}), positivity is the only one partly checkable from the data.
It requires \(0 < g(X) < 1\), where \(g(X)=P(A=1\mid X)\) is the propensity score, so it is a statement about the \emph{observed} distribution of treatment given the covariates.
That is why it comes first: before choosing between AIPW, TMLE, or DML, check whether the data can support any of them.

\subsection{What positivity is, and why it decides
feasibility}\label{what-positivity-is-and-why-it-decides-feasibility}

Positivity --- also called overlap or common support --- asks a simple
question: for every kind of person in the study, could they plausibly
have been either treated or untreated? In NHEFS the treatment \(A\) is
quitting smoking, and we model \(g(X)=P(\text{quit}\mid X)\) (the
companion code). The diagnostic is then whether quitters and
non-quitters occupy the same region of estimated propensity. Where they
do, each treated person has comparable untreated people to borrow an
outcome from, and the estimator interpolates. Where they do not --- a
covariate stratum that is essentially all-treated or all-untreated ---
there is nothing to compare against, and any number reported there is
extrapolation from the outcome model, not evidence from data. No amount
of estimator sophistication manufactures the missing comparison. This is
the organizing claim of the tutorial: overlap, not the estimator label,
decides whether the question is answerable.

\subsection{Reading the NHEFS overlap
diagnostic}\label{reading-the-nhefs-overlap-diagnostic}

\begin{figure}[htbp]\centering
\includegraphics[width=0.90\linewidth]{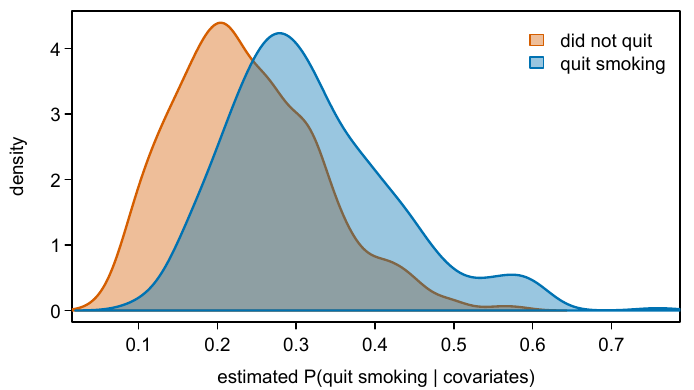}
\caption{\label{fig:nhefs-overlap}NHEFS: estimated propensity densities by exposure group. The fitted scores range from $0.044$ to $0.759$, with substantial overlap between the groups.}
\end{figure}

Figure~\ref{fig:nhefs-overlap} shows the NHEFS diagnostic: smoothed estimated-propensity densities, one curve per exposure group.
The two distributions sit almost on top of each other, and both lie within \([0.044,\,0.759]\).
The quitters are shifted only modestly to the right, and neither curve piles up against \(0\) or \(1\).
This is good overlap on the estimated-propensity scale: neither group sits where the other is effectively absent.

A one-dimensional score plot cannot show that every region of covariate space contains both groups, so it is reassuring rather than conclusive.
The diagnostic reads the \emph{estimated} propensity from its own Super Learner fit, a three-learner subset of the Step~2 library: logistic regression, lasso, and random forest (the companion code).
It therefore assesses practical overlap rather than proving structural positivity.
In borderline cases, also inspect covariate balance and the tails of the estimated weights.

Practical overlap is what the sample at hand and its fitted propensities show.
Structural positivity is the population condition (A3) of Box~\ref{box:b2}, fixed by the treatment-assignment mechanism.
The caution runs both ways: poor practical overlap does not by itself establish a structural violation \citep{petersen2012}.
Chance sparsity or an overfit propensity model can also produce it.
What poor practical overlap does flag is a risk that estimates using those fitted propensities are unstable, to be checked through the weights and sensitivity analyses.

Report three things, every time:

\begin{enumerate}
\def\labelenumi{\arabic{enumi}.}
\tightlist
\item
  the propensity distribution by exposure group --- histograms or
  smoothed densities as in Figure \ref{fig:nhefs-overlap} --- so a
  reader can \emph{see} the overlap rather than take it on trust;
\item
  the fraction of observations whose estimated propensity falls outside
  a pre-specified truncation window, which quantifies how much of the
  sample sits in the near-violation region;
\item
  the truncation rule you applied, and why.
\end{enumerate}

Table \ref{tab:nhefs-overlap} collects these for NHEFS.

\begin{table}[!ht]
\centering
\caption{NHEFS overlap diagnostics for the fitted propensity model $\hat g(X)=\hat P(A=1\mid X)$. Every estimated score lies inside the truncation window, so the bounds bind on nothing here.}
\label{tab:nhefs-overlap}
\begin{tabular}{lr}
\toprule
Diagnostic & Value \\
\midrule
Sample size $n$ & 1566 \\
Quit smoking ($A=1$) & 403 (25.7\%) \\
Estimated propensity range & 0.044--0.759 \\
Truncation window & [0.025,\,0.975] \\
Observations with capped propensity & 0 (0.0\%) \\
\bottomrule
\end{tabular}
\end{table}

\subsection{Truncation, and why weak overlap hurts every route}\label{truncation-and-why-weak-overlap-hurts-every-route-equally}

We truncate the estimated propensity to \([0.025,\,0.975]\) throughout,
following \citet{petersen2012}. This caps each inverse-probability weight at \(40\). It does not guarantee balanced relative influence or prevent a few observations from dominating the estimate
(Box~\ref{box:b8}).

Because NHEFS has good overlap, truncation is a safeguard rather than an active intervention: every estimated score already lies inside \([0.044,\,0.759]\), so zero of the \(1566\) observations are affected (Table~\ref{tab:nhefs-overlap}).
The same holds for the five-learner analysis fits we audited: the full-sample fit, one five-fold cross-fit, and all \(101\) double-cross-fit runs (companion code).
The point estimates from these fits are therefore unaffected by the rule (Box~\ref{box:b3}).

\begin{biobox}[label={box:b3}]
Strong positivity means that $\varepsilon\le g(X)\le1-\varepsilon$ almost surely for some $\varepsilon>0$.
Its bite is visible directly in the efficient influence function through the clever covariate $H(A,X)=A/g(X)-(1-A)/\{1-g(X)\}$.
The semiparametric efficiency bound is the smallest asymptotic variance of $\sqrt n(\hat\psi-\psi)$ that any regular asymptotically linear estimator of $\psi$ can attain.
It equals
$$
\operatorname{Var}\!\big(D(O)\big)=E\!\left[\frac{\sigma^{2}(1,X)}{g(X)}+\frac{\sigma^{2}(0,X)}{1-g(X)}\right]+\operatorname{Var}\!\big(Q_{1}(X)-Q_{0}(X)\big),
$$
where $\sigma^{2}(a,X)=\operatorname{Var}(Y\mid A=a,X)$.
Propensities near $0$ or $1$ inflate the first term.
Whether the bound stays finite depends on whether $\sigma^{2}(1,X)/g(X)$ and $\sigma^{2}(0,X)/\{1-g(X)\}$ are integrable and $Q_{1}(X)-Q_{0}(X)$ is square-integrable.

Propensities approaching $0$ or $1$ do not by themselves make the bound infinite.
For example, let $X$ be uniform on $(0,1)$, set $g(X)=\sqrt{X}/2$, and assume unit conditional variances and a constant conditional treatment effect.
Then $g$ approaches $0$, yet $E\{1/g(X)\}=4$ and the bound is finite.
Where the bound is large, it caps the achievable precision of AIPW, TMLE, and DML alike.

Weak overlap also enters the estimation error.
Box~\ref{box:b6} gives the error expansion and defines its second-order remainder, $R_2$, which obeys $|R_2|\lesssim\lVert \hat g-g\rVert\,\lVert \hat Q-Q\rVert$.
The constant in this bound is of order $\sup_x 1/\{\hat g(x)(1-\hat g(x))\}$, the estimated counterpart of $1/\{g(1-g)\}$; Box~\ref{box:b6} also gives the exact bound.
Weak overlap therefore magnifies both the leading variance and the second-order remainder.

The estimators share these variance and remainder bounds.
Their finite-sample behaviour under weak overlap can differ (Step~6).
In NHEFS the $[0.025,0.975]$ truncation is inactive, both for the overlap diagnostic (Table~\ref{tab:nhefs-overlap}) and for the analysis fits we audited (above).
Clipping therefore leaves the estimator unchanged, and it still targets the ATE.
\end{biobox}

\subsection{When the diagnostic fails: a
preview}\label{when-the-diagnostic-fails-a-preview}

Good overlap is not guaranteed, and the first move is worth just as much
when it returns bad news. In Step~6 we repeat exactly this diagnostic on
the LaLonde-PSID job-training data. There the two propensity densities barely overlap.
Enforcing common support by trimming to the overlap region discards \(84\)--\(89\%\) of the sample, and what survives can no longer pin down the target.
The diagnostic's role there is to say, up front, that no estimator, whether AIPW, TMLE, or DML, can recover an answer the data do not contain.
Diagnosing overlap first is what separates that verdict from a falsely precise number.

NHEFS shows substantial overlap on the estimated-propensity scale.
As the estimator panel in Step~4 will show, the doubly-robust estimates agree to within \(0.07\) kg of one another once the split is repeated, and every doubly-robust interval covers the canonical \(\sim\)3.4--3.5 kg benchmark \citep{hernan2020}.
With overlap examined, the remainder of the tutorial concentrates on \emph{why} AIPW, TMLE, and DML solve one estimating equation.

\textbf{Key takeaway.} Diagnose overlap first: where the treatment
groups do not overlap on the confounders, the data alone cannot pin down the effect, and no estimator recovers it without additional modelling assumptions.

\begin{tcolorbox}[title=Step~1 in pseudocode --- diagnose overlap first, colback=blue!3, colframe=blue!35, fonttitle=\bfseries, boxrule=0.4pt, arc=1pt]
$\hat g(X) \leftarrow$ Super Learner for $P(A{=}1\mid X)$ \\
plot density of $\hat g$ by treatment group; \\
report range, \% outside $[0.025, 0.975]$, truncation rule \\
\textit{\# if the groups do not overlap} $\rightarrow$ \textit{STOP: no stable estimate of $\psi$ without extrapolation}
\end{tcolorbox}

\section{Step~2 --- One shared nuisance
engine}\label{step-2-one-shared-nuisance-engine}

Having examined overlap, we now build the nuisance-estimation engine shared by the routes. Every doubly-robust estimator in this
tutorial reads from the same two learned quantities: the two
\emph{nuisance} functions, which we must estimate en route to \(\psi\)
but do not care about for their own sake. These are the propensity score
\(g(X)=P(A=1\mid X)\) and the outcome regression
\(Q(a,X)=E[Y\mid A=a,X]\), whose two arms we write \(Q_1\) and \(Q_0\).

AIPW adds the mean propensity-weighted residual correction to the outcome-regression plug-in.
TMLE instead updates the initial outcome fit through a logistic fluctuation.
For the ATE, DML uses the AIPW combination with out-of-fold nuisance fits (Step~4).

To isolate the combination rule, we give the one-step (AIPW) and targeted (TMLE) estimators identical initial nuisance predictions within each fitting scheme.
This step constructs these inputs; Steps~3--4 show how the estimators combine them.

\subsection{A single Super Learner for both
nuisances}\label{a-single-super-learner-for-both-nuisances}

We fit \(g\) and \(Q\) with the same ensemble --- the \textbf{Super
Learner} \citep{vanderlaan2007, polley2024superlearner}. A Super Learner
is not a model; it is a recipe for combining several candidate models.
You hand it a \emph{library} of learners, it scores each by
cross-validation, and it combines them with weights estimated from their cross-validated predictions. No single learner in the library needs to be correct.
A flexible learner is therefore insurance against having guessed the functional form wrong.
Appendix~\ref{app:superlearner} states the finite-grid oracle result and its conditions.
Our engine uses normalized non-negative least-squares (NNLS) weights, for which that result does not directly provide a guarantee (Box~\ref{box:b4}).

Our library holds five learners, chosen to span ``simple and stable''
through ``flexible and nonlinear'':

\begin{itemize}
\tightlist
\item
  \texttt{SL.mean} --- the sample mean; a floor that anchors the
  ensemble when the covariates carry little signal.
\item
  \texttt{SL.glm} --- main-effects linear/logistic regression; the
  classical parametric workhorse.
\item
  \texttt{SL.glmnet} --- lasso-penalized regression (the \texttt{SL.glmnet} default); shrinks coefficients and selects among many correlated covariates.
\item
  \texttt{SL.ranger} --- random forest; captures interactions and
  nonlinearity without being told where they are.
\item
  \texttt{SL.earth} --- multivariate adaptive regression splines; smooth
  nonlinear surfaces.
\end{itemize}

The \emph{same} five-learner library fits both nuisances, so \(\hat g\)
and \(\hat Q\) are built to a common standard rather than one being
tuned and the other left crude (the companion code).

\textbf{Choosing the library and the metalearner.} Two settings define a
Super Learner: the \emph{library} of candidate learners and the
\emph{metalearner} that combines them. We use the default convex metalearner rather than the \emph{discrete} Super Learner, which instead keeps only the single cross-validation-best learner.
The convex metalearner fits non-negative coefficients by NNLS and then normalizes them to sum to one (Appendix~\ref{app:superlearner}).
What matters for the library is diversity of functional form: a stable parametric anchor, a penalized learner for many correlated covariates, and at least one flexible nonparametric learner, so that no single modelling assumption governs the fit.
In high dimensions, pair the learners with covariate screeners. Five learners spanning these classes kept the computation manageable here
\citep{phillips2023, polley2024superlearner}.

\textbf{Key takeaway.} One shared Super Learner library feeds every estimator. Within each fitting scheme, the one-step and targeted estimators also share the same folds and initial nuisance fits, so any difference between them is the estimator's mechanics, not a luckier nuisance fit.

\begin{tcolorbox}[title=Step~2 in pseudocode --- one shared nuisance engine, colback=blue!3, colframe=blue!35, fonttitle=\bfseries, boxrule=0.4pt, arc=1pt]
\texttt{for} learner \texttt{in} \{\texttt{SL.mean, SL.glm, SL.glmnet, SL.ranger, SL.earth}\}: \\
\hspace*{1em}fit, score by $V$-fold CV \\
$\hat g(X) \leftarrow$ Super Learner for $P(A{=}1\mid X)$; \qquad $\hat Q(a,X) \leftarrow$ Super Learner for $E[Y\mid A,X]$ \\
\textit{\# Within each fitting scheme, the one-step and targeted estimators share the same initial nuisance fits.}
\end{tcolorbox}

\subsection{Two knobs that shape the nuisance fits}\label{the-two-knobs-that-actually-move-the-answer}

We examine two principal settings of the shared engine, the \textbf{learner library} and the \textbf{cross-fitting fold scheme}, which shape the nuisance fits and, through them, the numbers that come out downstream.
They are not the only influences: in any single run, the realized fold partition and the learners' internal randomness (random-forest draws, Super Learner's internal cross-validation folds) also move the fits (Steps~5 and~7).
A third knob, the propensity-truncation bound, is inert in the audited NHEFS fits, where it caps no estimated propensity (Step~1).
Step~6 examines propensity-based trimming, which changes the retained population rather than merely capping weights.
The estimator label (AIPW vs.~TMLE) is not one of these choices.
On NHEFS it barely moves the estimate, although with in-sample flexible nuisances it can matter (RHC, Step~4).

Table~\ref{tab:knobs} makes the ranking concrete on the NHEFS analysis (effect of quitting smoking on 1971--1982 weight change, \(n=1566\), \(403\) quitters).
Its entries are descriptive contrasts on this single dataset rather than a fully crossed factorial, but the ordering is stark.
Holding the library and folds fixed, swapping AIPW for TMLE moves the estimate by at most \(0.01\) kg.
At full sample, AIPW gives \(3.32\) and TMLE \(3.33\).
At the repeated five-fold cross-fit (\(r=100\)), DML gives \(3.38\) and CV-TMLE \(3.39\).
At the repeated three-fold double cross-fit, DC-AIPW gives \(3.37\) and DC-TMLE \(3.36\).

Changing the fold scheme (full-sample, five-fold cross-fit, or three-fold double-cross-fit) moves the estimate by up to \(0.07\) kg.
Varying the label and the scheme together, the six doubly-robust estimates span \(3.32\)--\(3.39\) kg.
Switching software implementations moves the estimate by up to \(0.18\) kg across single runs.
That comparison, however, changes several settings at once and does not isolate the library (Step~7).
Of the settings compared in Table~\ref{tab:knobs}, the one most often argued over, the estimator label, matters least.

\begin{table}[!ht]
\centering
\caption{What moves the NHEFS estimate and what does not (effect of smoking cessation on 1971--1982 weight change, kg). The first three rows compare estimates within our engine, which shares the five-learner Super Learner library and the $[0.025,0.975]$ propensity truncation; the bound clips no propensity in the audited fits (Step~1). Full-sample estimates are single fits, and split-based estimates are medians over $r=100$ splits. With the folds also held fixed, swapping the estimator label barely perturbs $\hat\psi$, whereas changing the fitting scheme moves it several times more. Varying both, the six doubly-robust estimates span $3.32$ kg (full-sample AIPW) to $3.39$ kg (CV-TMLE). The last row compares single runs of five software implementations of the same estimand; they span $3.32$--$3.49$ kg and differ in several settings at once (Step~7). Every spread is a fraction of the confidence half-width ($\approx 0.9$ kg).}
\label{tab:knobs}
\begin{tabular}{L{0.42\linewidth}lr}
\toprule
Knob varied & Held fixed & Spread in $\hat\psi$ (kg) \\
\midrule
Estimator label (AIPW vs.\ TMLE) & library, folds, truncation & $\le 0.01$ \\
Cross-fitting scheme & label, library, truncation & $0.07$ \\
Label and scheme together (six estimates) & library, truncation & $0.07$ \\
Whole software implementation (single runs) & estimand only & $0.18$ \\
\bottomrule
\end{tabular}
\end{table}

The fold scheme can matter because flexible learners can overfit their own
training rows. Fitting \(\hat g\) and \(\hat Q\) on the full sample and
then evaluating the influence function on those same rows ties each fit to the rows it scores; splitting into folds and predicting out of fold
breaks that tie \citep{chernozhukov2018, zivich2021}. NHEFS has no ground truth, so the shift between schemes does not show which scheme is closer to the true effect.

We use \(V=5\) outer folds for the cross-fit estimators (DML, CV-TMLE) and \(V=3\) splits for the double-cross-fit variants, which place \(g\) and \(Q\) on disjoint folds.
A single fold split (\(r=1\)) is the base configuration.
The panels also report the repeated-split median (\(r\)~up to \(100\), Step~5).
The choice of the number of splits and repetitions is studied by \citet{karim2025splits}; see also \citet{newey2018}, \citet{zivich2021} and \citet{mondol2025}.
Internally, each Super Learner picks its ensemble weights by its own
five-fold cross-validation. All propensity scores are truncated to
\([0.025,\,0.975]\) throughout, so truncation is held constant and
cannot masquerade as an estimator effect.

\subsection{Fair by construction}\label{fair-by-construction}

Within each fitting scheme, the one-step and targeted estimators use the same initial nuisance predictions, folds and truncation.
Their comparison therefore isolates the combination rule.
Comparing different fitting schemes also changes the nuisance fits; Table~\ref{tab:knobs} reports those empirical differences, and Box~\ref{box:b4} gives the technical background.

\begin{biobox}[label={box:b4}]
Our Super Learner combines the library learners with the \texttt{SuperLearner} package's default weights: non-negative least-squares (NNLS) coefficients, normalized to sum to one unless all are zero.
These weights need not minimize cross-validated risk over all convex combinations.
Appendix~\ref{app:superlearner} defines the weight selectors and states the finite-grid oracle result, which concerns a different selector and does not directly establish a guarantee for these NNLS weights.
Good prediction risk alone also does not deliver the nuisance rates that inference on $\psi$ needs (Box~\ref{box:b11}).

Within each fitting scheme, the one-step and targeted estimators of Steps~3--4 share the initial fits $\hat g$ and $\hat Q$; when cross-fit, these come from the same partition.
Their second-order remainders $R_2$ (Box~\ref{box:b6}) have the same product form but are evaluated at different outcome fits.
The one-step remainder satisfies $|R_2|\lesssim \lVert \hat g - g\rVert\,\lVert \hat Q - Q\rVert$.
TMLE's remainder bound replaces the initial $\hat Q$ by the targeted update $\hat Q^{*}$ (Box~\ref{box:b8}), and its empirical-process term differs in the same way.
For TMLE, the product-rate and empirical-process conditions therefore apply to the targeted fit $\hat Q^{*}$ (Box~\ref{box:b11}).
When those conditions hold, the two estimators share the leading term $\mathbb{P}_n D$, and any remaining difference between them is a finite-sample effect of how each solves that one score.
Steps~3--4 isolate this comparison.

Changing the learner library, the software, the fitting scheme or the random split instead changes the nuisance fits themselves.
Among fits meeting the conditions of Box~\ref{box:b11}, changes from the combination rule and changes from the nuisance fits are both $o_P(n^{-1/2})$, so the theory does not rank them.
The ranking is empirical: on NHEFS the fitting-scheme and software shifts in Table~\ref{tab:knobs} are several times the label shift, and all stay well inside the sampling half-width.
\end{biobox}

\section{Step~3 --- The shared efficient influence
function}\label{step-3-the-shared-efficient-influence-function}

With the data assembled and the confounder set fixed (Steps~0--2), we
can now write down the one quantity that AIPW, TMLE, and double machine
learning all target. This is the conceptual heart of the tutorial: once
you see that all three estimators are trying to solve the \emph{same}
equation, the differences between them stop looking like a menu of
competing methods and start looking like three routes to one answer.
Table~\ref{tab:notation} collects the notation and terminology used
throughout.

\begin{table}[!ht]
\centering
\caption{Notation and terminology used throughout this tutorial.}
\label{tab:notation}
\begin{tabular}{@{}L{0.30\linewidth}L{0.62\linewidth}@{}}
\toprule
Symbol / term & Meaning \\
\midrule
$\psi = E[Y^{1}-Y^{0}]$ & average treatment effect (ATE); the target estimand \\
$A,\ Y,\ X$ & treatment (0/1), outcome, measured confounders \\
$g(X)$ & propensity score $P(A{=}1\mid X)$ (a \emph{nuisance}) \\
$Q(a,X)$ & outcome regression $E[Y\mid A{=}a,X]$ (a nuisance); arms $Q_{1},Q_{0}$ \\
$D(O)$ & efficient influence function (EIF) of $\psi$ \\
$H(A,X)$ & clever covariate, $A/g(X)-(1-A)/\{1-g(X)\}$ \\
Super Learner & cross-validated ensemble of candidate learners \\
Cross-fitting & out-of-fold nuisance estimation ($V$ folds) \\
Double cross-fitting & $g$ and $Q$ fit on disjoint folds \\
AIPW\,/\,one-step & plug-in $+$ empirical mean of the EIF correction \\
TMLE & targeted fluctuation of $Q$ until the EIF term is zero \\
DML & cross-fit AIPW for the ATE (Neyman-orthogonal moment; Box~\ref{box:dmlaipw}) \\
\bottomrule
\end{tabular}
\end{table}

\subsection{The estimand and the
g-formula}\label{the-estimand-and-the-g-formula}

Recall from Step~0 the target: the average treatment effect (ATE) of
quitting smoking (\(A=1\)) versus not quitting (\(A=0\)) on 1971--1982
weight change \(Y\) (kg), \[
\psi \;=\; E\!\left[Y^{1} - Y^{0}\right],
\] where \(Y^{a}\) is the weight change we \emph{would} have seen had a
person's smoking status been set to \(a\). We never see both potential outcomes for anyone, so \(\psi\) is not directly a sample average. Under the identification assumptions in Box~\ref{box:b2} (Step~0), \(\psi\) is identified from the observed data \(O=(X,A,Y)\) by the
\textbf{g-formula} \[
\psi \;=\; E\!\left[\,Q_{1}(X) - Q_{0}(X)\,\right], \qquad Q_{a}(X) = E[Y \mid A=a, X].
\] Everything downstream is a way of estimating this contrast well
\citep{hernan2020}.

\subsection{Two nuisances, two singly-robust
baselines}\label{two-nuisances-two-singly-robust-baselines}

We use two \textbf{nuisance} functions to construct the estimators:

\begin{itemize}
\tightlist
\item
  the \textbf{propensity} \(g(X)=P(A=1\mid X)\) --- how likely each
  person was to quit, given their covariates;
\item
  the \textbf{outcome regression} \(Q(a,X)=E[Y\mid A=a,X]\) --- the
  expected weight change at treatment level \(a\); write
  \(Q_{1}=Q(1,X)\), \(Q_{0}=Q(0,X)\).
\end{itemize}

Each nuisance on its own gives a classical estimator, and each is
\textbf{singly robust}~--- correct only if \emph{its} model is correct.

\textbf{g-computation} fits the outcome model, predicts both potential
outcomes for everyone, and averages the contrast:
\(\hat\psi_{\mathrm{g}} = \tfrac1n\sum_{i}\{\hat Q_{1}(X_{i})-\hat Q_{0}(X_{i})\}\).
Its consistency relies on \(\hat Q\) being right. On NHEFS this returns
\(3.27\) kg (Table \ref{tab:eif-repair}).

\textbf{IPTW} (inverse-probability-of-treatment weighting) fits the
propensity model \(g\) instead and reweights the observed outcomes so the
treated and untreated groups resemble the whole population; in
stabilized (Hájek) form, \[
\hat\psi_{\mathrm{IPTW}} \;=\; \frac{\sum_{i} A_{i}Y_{i}/\hat g_{i}}{\sum_{i} A_{i}/\hat g_{i}} \;-\; \frac{\sum_{i} (1-A_{i})Y_{i}/(1-\hat g_{i})}{\sum_{i} (1-A_{i})/(1-\hat g_{i})}.
\] Its consistency relies on \(\hat g\) being right; it returns \(3.19\)
kg. Both baselines use the shared Super Learner nuisances, with
\(\hat g\) truncated to \([0.025, 0.975]\), which caps each
inverse-probability weight at \(40\) (the companion code).

These two numbers sit close on NHEFS, but each rests on a single
modelling bet. The efficient influence function is the object that lets
us bet on \emph{both} at once and lose only if \emph{both} are wrong.

\subsection{The efficient influence function, term by
term}\label{the-efficient-influence-function-term-by-term}

For \(\psi\) in the nonparametric model there is one canonical gradient,
the \textbf{efficient influence function} (EIF): \[
D(O) \;=\; \underbrace{H(A,X)\,\bigl(Y - Q_{A}(X)\bigr)}_{\text{propensity-weighted residual correction}} \;+\; \underbrace{\bigl(Q_{1}(X)-Q_{0}(X)\bigr)}_{\text{outcome contrast (g-formula plug-in)}} \;-\; \psi,
\] built from the \textbf{clever covariate} \[
H(A,X) \;=\; \frac{A}{g(X)} - \frac{1-A}{1-g(X)}.
\] This is the single object every later step refers back to.
Intuitively, an influence function measures how much each observation
pulls the estimate; the \emph{efficient} one has the smallest possible
variance, so no regular estimator of \(\psi\) can have a smaller asymptotic variance than \(\operatorname{Var}(D)/n\). Read left to right, the EIF is the
g-computation contrast plus a correction:

\begin{itemize}
\tightlist
\item
  The \textbf{outcome contrast} \(Q_{1}-Q_{0}\) is exactly the g-formula
  integrand --- the model-based guess.
\item
  The \textbf{residual} \(Y - Q_{A}\) is how wrong that guess was for
  the treatment level a person actually received. If \(\hat Q\) were
  perfect, these residuals would average to noise.
\item
  The \textbf{clever covariate} \(H\) weights each residual by inverse
  propensity, with a \(+\) sign for the treated and a \(-\) sign for the
  untreated. It routes each person's leftover error back to the arm they
  belong to, scaled up precisely where a covariate profile was
  under-represented in that arm. That weighting is what makes the
  correction do its job, which is why \(H\) carries the name \emph{clever}.
\end{itemize}

So the correction term \(H\,(Y-Q_{A})\) is the IPTW idea
(inverse-propensity weighting) applied not to the raw outcome but to the
\emph{residual} the outcome model left behind. The \textbf{AIPW /
one-step} estimator adds the empirical mean of this correction to the
plug-in, \[
\hat\psi_{\mathrm{AIPW}} \;=\; \frac1n\sum_{i}\Bigl\{\bigl(\hat Q_{1}(X_{i})-\hat Q_{0}(X_{i})\bigr) + \hat H(A_{i},X_{i})\bigl(Y_{i}-\hat Q_{A_{i}}(X_{i})\bigr)\Bigr\},
\] with \(\hat H\) the clever covariate evaluated at \(\hat g\). It solves the estimating equation \(\mathbb{P}_{n} D = 0\) in a
single step and carries an analytic standard error read straight off the
influence curve, with no bootstrap (the companion code). On NHEFS it
returns \(3.32\) kg (95\% CI \(2.50\),~\(4.14\)).
TMLE, which solves the
same equation by targeting rather than adding (Step~4), returns \(3.33\)
kg (95\% CI \(2.50\),~\(4.15\)).
The doubly-robust correction pulls the
two singly-robust baselines (\(3.27\) and \(3.19\) kg) onto essentially
one value, \(3.32\)--\(3.33\) kg.
It also brings both efficient estimators together, and both intervals cover the canonical Hernán--Robins \(\sim\)3.4--3.5 kg benchmark \citep{hernan2020}.

\begin{table}[t]
\centering
\caption{Singly-robust baselines and the doubly-robust repair on NHEFS (effect of smoking cessation on 1971--1982 weight change; $n=1566$, $403$ quitters, good overlap). The two singly-robust estimators each require one working model; the efficient-influence-function estimators require only that \emph{one of the two} be correct, and coincide here. Baselines are reported as point estimates: an analytic interval would need each baseline's working model to be a correctly specified parametric model (Box~\ref{box:b15}). The ``Must be correct'' column is the condition for point-estimator \emph{consistency}, not for valid inference. With flexible learners, the full-sample AIPW/TMLE intervals are reference Wald intervals, and inference relies on the cross-fit estimators (Step~5).}
\label{tab:eif-repair}
\begin{tabular}{llrr}
\toprule
Estimator & Must be correct (for consistency) & Estimate (kg) & 95\% CI \\
\midrule
g-computation      & outcome model $Q$      & 3.27 & --- \\
IPTW (stabilized)  & propensity model $g$   & 3.19 & --- \\
AIPW (EIF one-step)& $Q$ \emph{or} $g$      & 3.32 & (2.50, 4.14) \\
TMLE (EIF targeting)& $Q$ \emph{or} $g$     & 3.33 & (2.50, 4.15) \\
\bottomrule
\end{tabular}
\end{table}

\begin{biobox}[label={box:b5}]
$D(O)$ is the canonical gradient of $\psi$ in the nonparametric model: it has mean zero at the truth, $E[D(O;g,Q,\psi)]=0$, and its variance is the semiparametric efficiency bound for $\psi$ \citep{robins1994,bickel1993,hines2022}. Every doubly-robust estimator in this tutorial is a way of solving the empirical analogue $\mathbb{P}_n D = 0$.
AIPW adds the mean correction; TMLE (Step~4) fluctuates $Q$ until the correction is exactly zero \citep{vanderlaan2006}.

Double robustness is a statement about the \emph{bias} incurred when the nuisances converge to wrong limits $g^\star, Q^\star$.
For the ATE, the asymptotic bias of the one-step estimator factorises:
$$
\begin{aligned}
\operatorname{plim}\hat\psi_{\mathrm{AIPW}} - \psi \;=\;& E\!\left[\frac{g^\star(X)-g(X)}{g^\star(X)}\bigl(Q_1^\star(X)-Q_1(X)\bigr)\right]\\
&+\; E\!\left[\frac{g^\star(X)-g(X)}{1-g^\star(X)}\bigl(Q_0^\star(X)-Q_0(X)\bigr)\right].
\end{aligned}
$$
Each summand is a \emph{product} of a propensity error $(g^\star-g)$ and an outcome-model error $(Q_a^\star-Q_a)$. It vanishes whenever either factor is zero, so the estimator is consistent if \emph{either} $g^\star=g$ or $Q^\star=Q$. That product structure, not any single model being right, is what ``doubly robust'' names.

The denominators are well defined because $0<g^\star(X)<1$ almost surely, as the $[0.025,0.975]$ truncation ensures.
If those bounds clip true propensity values, however, even a consistent propensity fit converges to the clipped $g$.
The propensity-correct case $g^\star=g$ therefore needs $g$ itself to lie within the bounds.
\end{biobox}

\subsection{Why the correction repairs both baselines: double
robustness}\label{why-the-correction-repairs-both-baselines-double-robustness}

Intuitively, the EIF hedges. g-computation trusts the outcome model;
IPTW trusts the propensity model; the correction term \(H(Y-Q_A)\) lets
each fix the other's mistakes. If the outcome model \(\hat Q\) is right,
the residuals \(Y-\hat Q_A\) have conditional mean zero and the whole
correction washes out. What remains is the (correct) g-formula plug-in, so a
wrong propensity model does no harm. If instead the propensity model
\(\hat g\) is right, the inverse-probability weighting inside \(H\) is
calibrated correctly and the correction removes whatever bias a wrong
\(\hat Q\) introduced. You lose only if \emph{both} models are wrong at
once. This is why the DR estimators in Table~\ref{tab:eif-repair} can be
trusted more than either baseline that feeds them, even though on this well-behaved dataset all four point estimates nearly coincide.

Double robustness protects against \emph{misspecification}. A second,
distinct property protects against \emph{estimation error} when the
nuisances are flexible machine-learning fits rather than fixed
parametric models. That property is what licenses double machine learning
(Box~\ref{box:b6}).

\begin{biobox}[label={box:b6}]
In DML terminology, the EIF is the moment
$D(O;g,Q,\psi)=(Q_1-Q_0)+H(A,X)(Y-Q_A)-\psi$.
It has mean zero at the truth.
It is also \textbf{Neyman-orthogonal}: the Gateaux derivative of its expectation in the nuisances is zero at the truth.

Consider admissible paths
$g_t=g+t(\bar g-g)$ and $Q_t=Q+t(\bar Q-Q)$.
Take $\bar g\in[c,1-c]$ for some $c>0$.
Then $g_t\in[tc,1-tc]$ for $0<t\le1$, while $g_0=g$ satisfies positivity (A3).
Assume that $D(O;g_t,Q_t,\psi)$ remains integrable near $t=0$.
For these directions,
$$
\left.\frac{\partial}{\partial t}\, E\!\left[D\bigl(O;g_t,Q_t,\psi\bigr)\right]\right|_{t=0} \;=\; 0 .
$$

The first-order effect of nuisance error on the estimating equation cancels, so nuisance error enters the estimation error only at second order:
$$
\hat\psi - \psi \;=\; \mathbb{P}_n D \;+\; (\mathbb{P}_n-P)\{\hat D-D\} \;+\; R_2, \qquad |R_2| \;\lesssim\; \bigl\lVert \hat g - g \bigr\rVert \cdot \bigl\lVert \hat Q - Q \bigr\rVert ,
$$
where $D=D(O;g,Q,\psi)$ is the true EIF and $\hat D=D(O;\hat g,\hat Q,\hat\psi)$ is its fitted version.
Under the conditions of Box~\ref{box:b11}, the empirical-process term $(\mathbb{P}_n-P)\{\hat D-D\}$ is $o_P(n^{-1/2})$, controlled by cross-fitting (Step~5) or by a Donsker condition.

For the one-step estimator, $R_2$ is exactly the bias product of Box~\ref{box:b5}, with $(\hat g,\hat Q)$ in place of $(g^\star,Q^\star)$,
$$
R_2=P\!\left[(\hat g-g)\left\{\frac{\hat Q_1-Q_1}{\hat g}+\frac{\hat Q_0-Q_0}{1-\hat g}\right\}\right],
$$
computed fold by fold and averaged with weights $n_k/n$ under cross-fitting.
If $\delta\le\hat g\le1-\delta$, the Cauchy--Schwarz inequality gives
\[
|R_2|\le\delta^{-1}\lVert\hat g-g\rVert\,
\bigl(\lVert\hat Q_1-Q_1\rVert+\lVert\hat Q_0-Q_0\rVert\bigr)
\]
in $L_2(P)$ norms.
The $[0.025,0.975]$ truncation makes $\delta^{-1}=40$.
So if each nuisance converges even at the slow rate $o_P(n^{-1/4})$, then $R_2=o_P(n^{-1/2})$, negligible next to the $O_P(n^{-1/2})$ leading term $\mathbb{P}_n D$.

Hence, under the conditions of Box~\ref{box:b11},
$\sqrt{n}(\hat\psi-\psi)\to N\!\bigl(0,\operatorname{Var}(D)\bigr)$,
with $\operatorname{Var}(\hat\psi)$ estimated by $\mathbb{P}_n\{\hat D^2\}/n$.

Cross-fitting fits $g$ and $Q$ on the folds other than the one being scored, which removes own-observation overfitting.
That is how it controls the empirical-process term for machine-learning nuisances without Donsker conditions \citep{chernozhukov2018}.
It leaves the product-rate condition on $R_2$ in place.
Double cross-fitting fits $g$ and $Q$ on disjoint folds to further break their dependence in the remainder \citep{newey2018}.
This orthogonality-plus-cross-fitting argument is precisely DML, applied to the same score AIPW already uses.
\end{biobox}

Double robustness and Neyman orthogonality are two faces of the same
product structure: one appears in the misspecification bias, the other in the
estimation remainder.
Both come free with the EIF.
That is why AIPW, TMLE, and DML are asymptotically equivalent and efficient once the nuisances satisfy the product-rate condition above and the empirical-process term is controlled, attaining variance
\(\operatorname{Var}(D)/n\) \citep{kennedy2023, hines2022, diaz2020}.
Step~4 presents these as three
routes to the single equation \(\mathbb{P}_n D = 0\) written above:
AIPW's one-step average, TMLE's targeting fluctuation, and DML's
out-of-fold evaluation.
Step~5 examines the conditions under which cross-fitting keeps the machine-learning version's inference valid.

\textbf{Key takeaway.} AIPW, TMLE, and DML all target one efficient
influence function; double robustness and Neyman orthogonality are two
faces of its product structure.

\begin{tcolorbox}[title=Step~3 in pseudocode --- the one shared score, colback=blue!3, colframe=blue!35, fonttitle=\bfseries, boxrule=0.4pt, arc=1pt]
$H(A,X) \leftarrow A/\hat g(X) - (1-A)/(1-\hat g(X))$ \quad \textit{\# clever covariate} \\
$\hat D(O) \leftarrow H\,(Y-\hat Q_A) + (\hat Q_1 - \hat Q_0) - \psi$ \quad \textit{\# efficient influence function} \\
baselines: g-comp $\psi \leftarrow \mathrm{mean}(\hat Q_1 - \hat Q_0)$; \quad IPTW $\psi \leftarrow$ H\'ajek-weighted mean of $Y$
\end{tcolorbox}

\section{Step~4 --- Two combination rules, three fitting schemes (one
score)}\label{step-4-two-combination-rules-three-fitting-schemes-one-score}

Steps~2 and 3 built the two ingredients every doubly-robust estimator
needs: the propensity score \(g(X)=P(A=1\mid X)\) (who quits smoking)
and the outcome regression \(Q(a,X)=E[Y\mid A=a,X]\) (the expected
1971--1982 weight change under each smoking status). A plug-in that
trusts only \(Q\) is g-computation; a plug-in that trusts only \(g\) is
IPTW. Each rests on one model being right. The doubly-robust estimators
splice the two. Under the usual regularity conditions, consistent estimation of \emph{either} nuisance suffices for consistency. The Wald inference used here is justified under the additional conditions of Box~\ref{box:b11}.

\begin{figure}[htbp]\centering
\includegraphics[width=\linewidth]{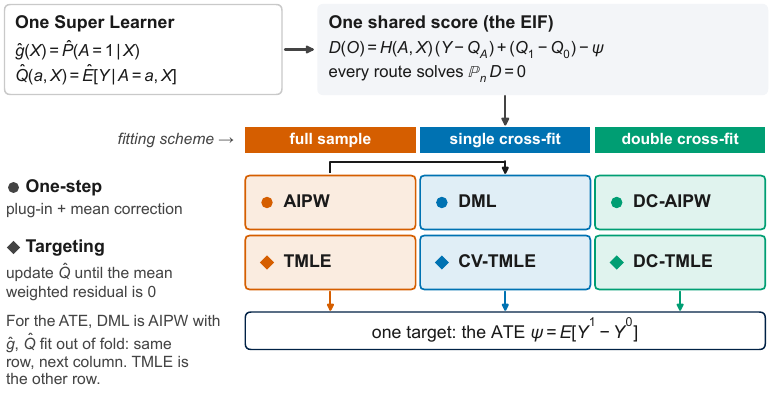}
\caption{\label{fig:concept}One ATE score, two combination rules and three fitting schemes. A shared nuisance engine supplies the propensity and outcome predictions. Rows distinguish adding a one-step correction from updating the outcome regression until the mean weighted residual is zero. Columns distinguish full-sample, single cross-fit and double cross-fit initial nuisance estimation. Colour identifies the fitting scheme; shape identifies the combination rule. For the ATE score and matched inputs, DML is cross-fit AIPW (Box~\ref{box:dmlaipw}). All six constructions target the same ATE and, under the conditions of Box~\ref{box:b11}, are asymptotically equivalent \citep{kennedy2023, hines2022, diaz2020}.}
\end{figure}

AIPW, TMLE, and DML are not three rival methods, and they do not even divide along a single axis.
All are built from the \emph{same} efficient influence function (EIF) of Step~3, and all solve the \emph{same} estimating equation \(\mathbb{P}_n D = 0\) (Figure~\ref{fig:concept}).
What differs sits on two separate axes.
The first is the \textbf{combination rule}, that is, how the EIF is turned into a number.
The \emph{one-step} rule adds the empirical mean of the EIF to the g-computation plug-in (the AIPW family).
\emph{Targeting} instead fluctuates \(\hat Q\) until that mean is exactly zero and then plugs in (the TMLE family).

The second axis is the \textbf{fitting scheme}: whether the nuisances \(\hat g,\hat Q\) are fit on the full sample, on a single cross-fit, or on disjoint (double) cross-fit folds (Figure~\ref{fig:fitting-schemes}).
Crossing the two axes gives a \(2\times3\) grid of six doubly-robust estimators (Figure~\ref{fig:concept}), not three parallel methods.
The point most easily missed is that \textbf{AIPW and DML share the same combination rule}.
DML \emph{is} the one-step estimator with the nuisances fit out of fold.
It changes \emph{how} \(\hat g,\hat Q\) are estimated, not the score they enter, so it is not a third way of solving the equation.
TMLE is the genuinely different move, the \emph{other} combination rule, though it lands on the same EIF and carries the same influence function.
We present three routes because the three literatures and their R packages ship them that way, and reading them side by side is what dissolves the confusion.

Under the conditions of Box~\ref{box:b11}, all six agree in large samples \citep{kennedy2023, hines2022, diaz2020}. In a finite sample their estimates can still differ materially, even with good overlap and shared initial nuisance fits. Targeting replaces the initial outcome fit \(\hat Q\) by the updated fit \(\hat Q^{*}\) before plugging in.
The targeted and one-step estimators therefore share the leading term of the error expansion in Box~\ref{box:b6} but not its remainder terms (Box~\ref{box:b4}).
The RHC subsection below shows such a gap when the nuisances are fit in sample, and Step~6 shows one on the LaLonde data, where overlap is poor.
What Figure~\ref{fig:nhefs-panel} adds is a concrete, reproducible instance of close agreement on real data.
On NHEFS, where overlap is good (Step~1), all six doubly-robust
estimators land within \(0.07\) kg of one another once the split is repeated. 

\subsection{Route 1 --- AIPW: the one-step rule, full
sample}\label{route-1-aipw-the-one-step-rule-full-sample}

The augmented inverse-probability-weighted (AIPW / AIPTW) estimator,
also called the one-step estimator, is the most literal reading of the EIF.
It is the top-left cell of the grid in Figure~\ref{fig:concept}: the one-step combination rule with full-sample
nuisances.
Start from the
g-computation plug-in \(\tfrac1n\sum_i\{\hat Q(1,X_i)-\hat Q(0,X_i)\}\)
and add the empirical mean of the propensity-weighted residual
correction. If the outcome model is off, the correction leans on the
propensity model to repair it, and vice versa. This is the doubly-robust
guarantee. AIPW is the classical construction of \citet{robins1994},
grounded in the semiparametric efficiency theory of \citet{bickel1993};
\citet{luque2018} gives an applied walk-through
(Box~\ref{box:b7}).

\begin{biobox}[label={box:b7}]
The one-step estimator adds the empirical mean of the estimated EIF to the plug-in:
$$\hat\psi_{\mathrm{AIPW}}=\frac1n\sum_{i=1}^{n}\Big[\hat Q(1,X_i)-\hat Q(0,X_i)+\hat H(A_i,X_i)\,\{Y_i-\hat Q(A_i,X_i)\}\Big].$$
Equivalently, $\hat\psi_{\mathrm{AIPW}}$ solves $\tfrac1n\sum_i \hat D(O_i)=0$ by construction, since the sample analogue of $E\{D(O)\}+\psi$ \emph{is} the estimator. Under the conditions of Box~\ref{box:b11} its influence function is $D(O)$ itself, giving the closed-form variance estimate $\widehat{\mathrm{Var}}(\hat\psi)=n^{-2}\sum_i \hat D(O_i)^2$ used for the Wald intervals in Figure \ref{fig:nhefs-panel}.
\end{biobox}

\subsection{Route 2 --- TMLE: the other combination rule, full
sample}\label{route-2-tmle-the-other-combination-rule-full-sample}

Targeted maximum likelihood estimation reaches the same score by the
\emph{other combination rule}: the second row of the grid in Figure~\ref{fig:concept}, still on the full sample.
Rather than add a correction to the plug-in \emph{after} the fact, TMLE updates the fitted outcome
model \(\hat Q\) by a small, propensity-informed amount and then reads
off an ordinary plug-in from the updated model.
The update is a one-parameter logistic ``fluctuation'' carrying the clever covariate
\(H\), chosen so that the EIF averages to \emph{exactly} zero.
Because the final answer is a substitution estimator evaluated on a
valid outcome model, it automatically respects the natural range of the parameter (for a risk difference, \([-1,1]\); Box~\ref{box:b10}), which matters when \(Y\) is bounded or a probability.
TMLE was introduced by \citet{vanderlaan2006}.
The modern machinery and its diagnostics are in \citet{gruber2010, gruber2012}, and
\citet{schuler2017} is a readable tutorial
(Box~\ref{box:b8}).

\begin{biobox}[label={box:b8}]
Rescale the outcome $Y$ to $[0,1]$ by its observed range.
This is the bounded-outcome device of \citet{gruber2010}, which treats the sample range as the outcome's bounds.
Then fit the logistic fluctuation submodel
$$\operatorname{logit}Q_{\epsilon}(a,X)=\operatorname{logit}\hat Q(a,X)+\epsilon\,H(a,X),$$
estimating $\hat\epsilon$ by logistic regression of the rescaled $Y$ with offset $\operatorname{logit}\hat Q$, covariate $H$, and no intercept.
Here $H$ is evaluated at the truncated $\hat g\in[0.025,0.975]$, and the rescaled $\hat Q$ is kept inside $(0,1)$ so that the offset is finite.

\textbf{Check existence.}
The logistic quasi-log-likelihood of this one-parameter fit is strictly concave in $\epsilon$.
Its score is
\[
S(\epsilon)=\sum_i H(A_i,X_i)\{Y_i-Q_{\epsilon}(A_i,X_i)\}.
\]
The score falls from a positive limit as $\epsilon\to-\infty$ to a negative limit as $\epsilon\to+\infty$, unless the $[0,1]$-rescaled outcome is perfectly separated along $H$.
Such separation means that every treated unit is at $Y=1$ and every control at $Y=0$, or the reverse.
No dataset here is separated in this way.
The score therefore has a single root, the unique finite maximizer $\hat\epsilon$ \citep[cf.][Theorem~3]{albert1984}.
Truncation is not what secures this.
Bounding $|H|\le 40$ bounds each observation's absolute contributions to the score and to the information.
It does not, however, guarantee balanced relative influence, and boundedness alone does not rule out separation.

\textbf{Check the numerical solution.}
The fitting algorithm finds that root only to its tolerance.
The code therefore checks convergence and records the empirical score residual.

\textbf{Compute the estimate.}
Update $\hat Q^{*}=Q_{\hat\epsilon}$ and take the plug-in
\[
\hat\psi_{\mathrm{TMLE}}=\frac1n\sum_i\{\hat Q^{*}(1,X_i)-\hat Q^{*}(0,X_i)\},
\]
mapped back to the original scale.
The first-order condition of that logistic fit is exactly
\[
\sum_i H(A_i,X_i)\{Y_i-\hat Q^{*}(A_i,X_i)\}=0,
\]
so the empirical EIF-mean is identically $0$ and $\hat\psi_{\mathrm{TMLE}}$ is a range-respecting substitution estimator.
Under the conditions of Box~\ref{box:b11}, applied to the targeted fit $\hat Q^{*}$, it has the same influence function $D(O)$ as AIPW.
\end{biobox}

\subsection{Route 3 --- DML: the one-step score again, fit out of
fold}\label{route-3-dml-the-one-step-score-again-fit-out-of-fold}

DML sits in AIPW's row of Figure~\ref{fig:concept}, one column to the right: each observation's \(\hat g\) and \(\hat Q\) come from models trained \emph{outside} that observation's fold (cross-fitting). This severs the feedback loop in which a flexible learner
fits its own training points more closely than new ones and so shrinks the residuals that the correction is built from (as on RHC, below). Cross-fitting is what licenses black-box learners for the
nuisances without restrictive complexity (Donsker) conditions
\citep{chernozhukov2018}. \citet{naimi2023} use simulations to examine machine learning
for the nuisances in epidemiology. The estimation library
used here is a Super Learner ensemble \citep{polley2024superlearner}.

The terminology needs pinning, because the same idea appears under several names.
Single cross-fitting of the AIPW score gives DML \citep{chernozhukov2018, diaz2020, qiu2024}.
CV-TMLE fits initial nuisances out of fold and pools the targeting fluctuation across held-out folds \citep{zheng2011}.
For each held-out fold, double cross-fitting fits the initial propensity and outcome regressions on two disjoint training folds; it gives DC-AIPW and DC-TMLE \citep{newey2018, zivich2021, mondol2025}.

\begin{figure}[htbp]\centering
\includegraphics[width=\linewidth]{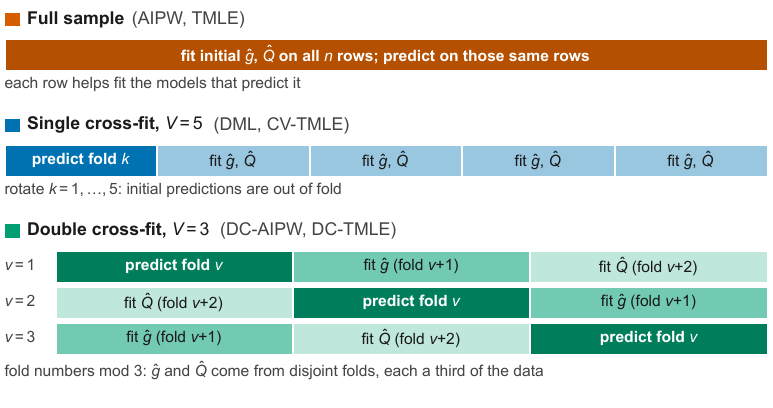}
\caption{\label{fig:fitting-schemes}Training and prediction folds of the initial nuisance fits under the three fitting schemes. Full-sample fitting (AIPW, TMLE) trains $\hat g$ and $\hat Q$ on all $n$ rows and predicts those same rows. Single cross-fitting with $V=5$ folds (DML, CV-TMLE) trains both on four folds and predicts the fifth, rotating over the folds. Double cross-fitting with $V=3$ folds (DC-AIPW, DC-TMLE) predicts held-out fold $v$ with $\hat g$ trained on fold $v+1$ and $\hat Q$ trained on fold $v+2$ (fold numbers mod 3); each of the three folds is shown once as the held-out fold. The diagram concerns the initial fits only: our CV-TMLE and DC-TMLE then pool one targeting fluctuation across the held-out folds. Colour identifies the fitting scheme, as in Figure~\ref{fig:concept}.}
\end{figure}

Figure~\ref{fig:fitting-schemes} shows which folds train the initial nuisance fits, and which fold they predict, under each fitting scheme.
We use one fixed rotation over three folds, as in the replication code of \citet{zivich2021}.
Each held-out fold is scored with $\hat g$ trained on the next fold and $\hat Q$ trained on the one after.
Our DC-TMLE pools one fluctuation across held-out folds, as our CV-TMLE does.
\citet{zivich2021} instead fit a separate fluctuation in each fold.

Because the fluctuation parameter is estimated, pooled and fold-specific targeting define distinct finite-sample estimators, even when the fluctuation is linear in that parameter.
Both solve the same score and, under the conditions of Box~\ref{box:b11}, share one influence function.
The analogous choice on the one-step side is innocuous, because AIPW fits no such parameter.
Cross-fit AIPW returns the same ATE estimate whether the per-fold scores are pooled or the per-fold estimates are averaged, exactly so for equal-sized folds \citep{chernozhukov2018} (Box~\ref{box:b9}).
DML and cross-fit AIPW denote one estimator when both use the ATE score and the same inputs.
The partially linear model that \citet[Sec.~4]{chernozhukov2018} treat first estimates a different quantity (Box~\ref{box:dmlaipw}).

\begin{biobox}[label={box:b9}]
Partition $\{1,\dots,n\}$ into $V$ folds. For fold $k$, fit $\hat g^{(-k)},\hat Q^{(-k)}$ on the complement and evaluate the score on fold $k$; average over folds, weighting fold $k$ by $n_k/n$.

The estimator has the error expansion of Box~\ref{box:b6}, with a leading EIF average, an empirical-process term and a second-order remainder $R_2$ computed fold by fold.
Under the consistency and positivity conditions of Box~\ref{box:b11}, cross-fitting makes the empirical-process term $o_P(n^{-1/2})$ without a Donsker condition.
The remainder is $o_P(n^{-1/2})$ if each nuisance converges at rate $o_P(n^{-1/4})$.
Double cross-fitting additionally estimates $g$ and $Q$ on disjoint training folds \citep{newey2018,zivich2021}.
\end{biobox}

\begin{biobox}[label={box:dmlaipw}]
For the ATE, DML and cross-fitted AIPW give exactly the same estimate when they use the same score and inputs.
This algebraic identity does not by itself guarantee valid inference.

\textbf{What is equal?}
DML is a general recipe: solve a Neyman-orthogonal score with cross-fitted nuisances \citep{chernozhukov2018}.
AIPW is one estimator built on one such score.
The AIPW formula can be evaluated with full-sample or cross-fitted nuisances.
In Figure~\ref{fig:concept}, the bare label AIPW denotes the full-sample version and DML the cross-fit version.

Write the moment of Box~\ref{box:b6} as $D(O;g,Q,\psi)=\varphi(O;g,Q)-\psi$, where
$\varphi=(Q_1-Q_0)+H(A,X)(Y-Q_A)$.
This is the ATE score of the interactive regression model \citep[eq.~5.3]{chernozhukov2018}, implemented as \texttt{DoubleMLIRM} \citep{bach2024}.
Let fold $k$ ($k=1,\dots,V$) hold the $n_k$ indices $I_k$, and let $\hat\eta_{-k}=(\hat g^{(-k)},\hat Q^{(-k)})$ be fit on its complement.
Pooled DML \citep[Def.~3.2]{chernozhukov2018} solves the cross-fit estimating equation
$$
\frac1n\sum_{k=1}^{V}\sum_{i\in I_k}\bigl\{\varphi(O_i;\hat\eta_{-k})-\psi\bigr\}=0
\quad\Longrightarrow\quad
\hat\psi_{\mathrm{DML}}=\frac1n\sum_{k=1}^{V}\sum_{i\in I_k}\varphi(O_i;\hat\eta_{-k}).
$$
The right-hand side is the cross-fit AIPW (one-step) estimator, term by term.

The implication holds because the score is linear in $\psi$ with slope $-1$.
Solving the estimating equation and adding the one-step correction are therefore the same operation.
For a parameter whose score is nonlinear in it, the two can give different estimators \citep{kennedy2023}.

\textbf{What must match?}
The identity requires the ATE score above and the same inputs.
Both estimators must use the same folds, the same out-of-fold nuisance predictions, the same truncation of $\hat g$, and the unnormalized weights in $H$.
Normalizing those weights, as the H\'ajek form does for the IPTW baseline, changes the estimator.
With repeated splits, both must also aggregate across splits by the same rule.

The same slope governs the standard errors.
The DML variance estimator scales the mean squared score by the inverse squared slope, so a slope of $-1$ makes it reduce to the $\mathbb{P}_n\{\hat D^2\}/n$ of Box~\ref{box:b6}.
For a single split with these inputs, the standard errors therefore agree too.
Software implementations need not match all of these inputs; on NHEFS, the spread among them is compatible with the fitting scheme plus single-split noise (Step~7).

\textbf{What does the identity establish?}
It establishes that two names denote one estimator, not that the estimator's inference is valid.
The identity is algebraic: it holds for any nuisance fits, accurate or not.
Valid inference is a separate matter and needs the rate and regularity conditions of Box~\ref{box:b11}.

Cross-fitting bears on that second question.
Full-sample AIPW solves the same equation with in-sample nuisances, so it relies on Donsker-type conditions, or another way to control the empirical-process term.
Cross-fitting controls that term without them, but it leaves the product-rate condition in place (Box~\ref{box:b9}).

TMLE and CV-TMLE solve the same equation by fluctuating $\hat Q$ (Box~\ref{box:b8}).
Targeting changes the outcome fit before the plug-in, so even with matched initial fits the targeted estimate need not equal the one-step estimate.
They are different estimators: under the conditions of Box~\ref{box:b11} they are asymptotically equivalent to AIPW and DML, but finite-sample agreement is not guaranteed.

\textbf{Why can partially linear DML differ?}
The partially linear model that \citet[Sec.~4]{chernozhukov2018} treat first uses a different score,
\[
\{Y-\ell(X)-\theta(A-g(X))\}\{A-g(X)\},
\qquad \ell(X)=E[Y\mid X].
\]
Its population root is
\[
\theta=\frac{E[v(X)\tau(X)]}{E[v(X)]},
\]
where $v(X)=\operatorname{Var}(A\mid X)=g(X)\{1-g(X)\}$ and $\tau(X)=Q_1(X)-Q_0(X)$.
This root is a variance-weighted effect.
It equals $\psi$ if and only if $\tau(X)$ is uncorrelated with $v(X)$, as under a constant effect.
DML with this score therefore targets $\psi$ only under that condition, whereas DML with the ATE score above is cross-fit AIPW.
\end{biobox}

\subsection{Agreement on NHEFS}\label{they-converge-on-nhefs}

\begin{figure}[htbp]\centering
\includegraphics[width=1.00\linewidth]{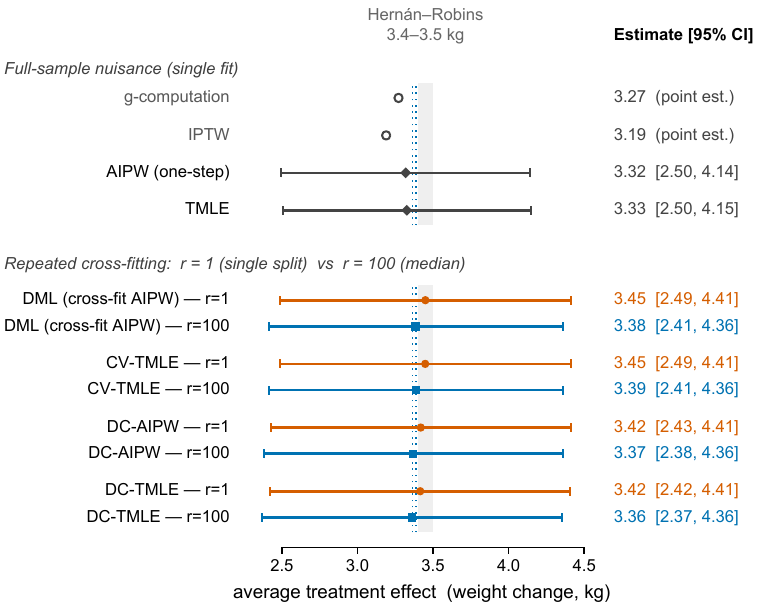}
\caption{\label{fig:nhefs-panel}NHEFS estimator panel: effect of smoking cessation on 1971--1982 weight change (kg), $n=1566$ (403 quitters). The panel shows the full-sample estimators (g-computation, IPTW, AIPW one-step, TMLE) and the four split-based estimators (DML, CV-TMLE, DC-AIPW, DC-TMLE). Each split-based estimator is shown both at a single split ($r=1$, orange) and median-aggregated over $r=100$ repeated cross-fits ($r=100$, blue). Wald intervals are not reported for the single-nuisance plug-ins (g-computation, IPTW). The grey band marks the canonical Hern\'an--Robins benchmark ($\sim$3.4--3.5 kg). The dotted lines mark the span of the four $r=100$ doubly-robust estimates. The six doubly-robust estimates span about $0.07$ kg at $r=100$ ($0.13$ kg with the single-split rows), and their intervals all cover the benchmark. Same Super Learner library, fold schemes, and $[0.025,0.975]$ truncation throughout. The full-sample AIPW and TMLE intervals are reference Wald intervals; inference with the flexible nuisance library relies on the cross-fit rows (Step~5).}
\end{figure}

Figure~\ref{fig:nhefs-panel} runs all eight estimators on the identical
estimand: the effect of smoking cessation on 1971--1982 weight
change, \(n=1566\) with \(403\) quitters (the companion code).
The two single-nuisance plug-ins (g-computation \(3.27\) kg, IPTW \(3.19\) kg)
sit alongside the six doubly-robust estimators.
These span from
\(3.32\)--\(3.33\) kg (full-sample AIPW and TMLE) up to \(3.39\) kg
(CV-TMLE at \(r=100\)), a spread of \(0.07\) kg (\(0.13\) kg if the single-split rows are included).
That spread is
more than an order of magnitude smaller than the width of any single
confidence interval (about \(1.6\) to \(2.0\) kg).
On this dataset, the choice among the six doubly-robust estimators is therefore immaterial next to
sampling uncertainty. Every doubly-robust interval covers the canonical
Hernán--Robins benchmark of roughly \(3.4\) to \(3.5\) kg for these data
\citep{hernan2020}.

Two patterns in Figure~\ref{fig:nhefs-panel} are worth reading, not as a ranking but as accounting.
First, the split-based estimates are a few hundredths of a kilogram higher than their full-sample counterparts.
At $r=100$, DML gives 3.38~kg, CV-TMLE 3.39~kg, DC-AIPW 3.37~kg, and DC-TMLE 3.36~kg.
Full-sample AIPW and TMLE give 3.32--3.33~kg.

Second, the split-based intervals are wider, already at a single split ($r=1$).
They are about 1.9~kg wide for the cross-fit routes and 2.0~kg for the double-cross-fit routes, compared with 1.6~kg for full-sample AIPW and TMLE.
For example, the DML interval at $r=100$ is $(2.41,4.36)$, versus $(2.50,4.14)$ for full-sample AIPW.

Behind the $r=100$ rows, the single-split estimates scatter with a standard deviation of about 0.08~kg for the cross-fit routes and 0.14~kg for the double-cross-fit routes.
On a single split, the six doubly-robust estimates span a median of 0.15~kg (interquartile range 0.09--0.22~kg; companion code).
These estimators differ in fitting scheme and combination rule; four of them also depend on the split drawn.

Repeating the split and median-aggregating moves each single-split ($r=1$) estimate down by 0.05--0.06~kg, the $r=1$-to-$r=100$ shift visible in the figure.
All six doubly-robust estimates then lie inside a 0.07-kg band.
Throughout, propensity scores were bounded to $[0.025,0.975]$ before weighting.

Good overlap, documented in Step~1, is a precondition for this convergence, not a guarantee of it. When positivity fails, the shared score
offers no rescue --- a case taken up in Step~6. The practitioner
corollary is the organizing thesis of the whole tutorial: diagnose
overlap first; only then does the equivalence of the routes become the
reason not to agonize over which one to run.

\subsection{The same machinery on a binary outcome:
RHC}\label{the-same-machinery-on-a-binary-outcome-rhc}

The NHEFS walkthrough exercises the three routes on a \emph{continuous} outcome, weight change in kilograms.
Applied analyses, though, more often target a \textbf{risk difference} for a \textbf{binary} outcome, so it is worth confirming that nothing in the one-EIF story is special to a mean on the real line.
We repeat the identical pipeline on the right-heart catheterization (RHC) data of \citet{connors1996}, the observational critical-care cohort that \citet{mondol2025} use to demonstrate DC-TMLE.
The pipeline keeps the same Super Learner library, the same fold schemes, and the same \([0.025,0.975]\) truncation.
Now the exposure \(A\) is receipt of a right-heart catheter, and the outcome \(Y\) is 30-day mortality.
The estimand is the ATE on the risk-difference scale, \(\psi = P(Y^{1}=1)-P(Y^{0}=1)\).
The sample is much larger (\(n = 5{,}735\)), and the confounder set is higher-dimensional (roughly sixty to seventy one-hot terms).
As with NHEFS, identification rests on no unmeasured confounding given the recorded baseline covariates, consistency, and positivity.

\begin{figure}[htbp]\centering
\includegraphics[width=1.00\linewidth]{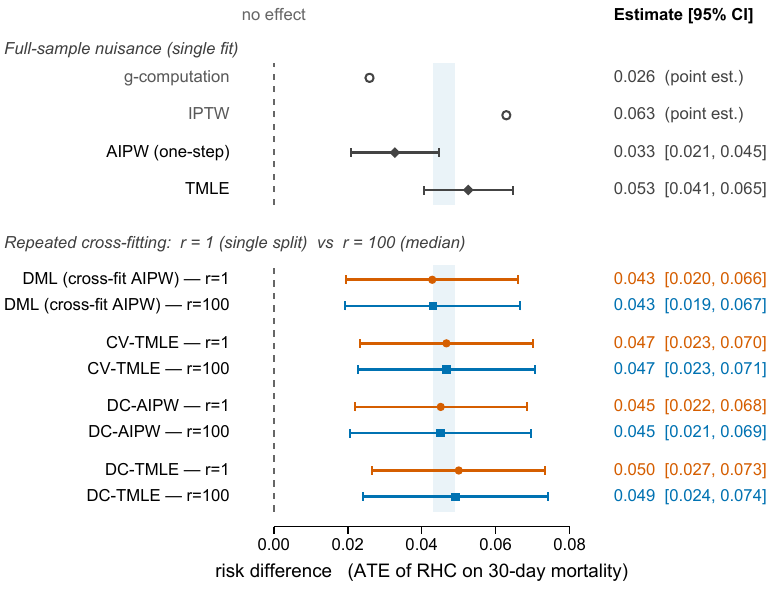}
\caption{\label{fig:rhc-panel}RHC estimator panel: effect of right-heart catheterization on 30-day mortality (risk difference), $n=5735$. Same Super Learner library, fold schemes, and $[0.025,0.975]$ truncation as the NHEFS panel (Figure~\ref{fig:nhefs-panel}), re-specialized to a binary outcome. The panel shows the full-sample estimators (g-computation, IPTW, AIPW one-step, TMLE) and the four split-based estimators (DML, CV-TMLE, DC-AIPW, DC-TMLE). Each split-based estimator is shown both at a single split ($r=1$, orange) and median-aggregated over $r=100$ repeated cross-fits ($r=100$, blue). The shaded band marks the span of the four $r=100$ doubly-robust estimates; g-computation and IPTW have no analytic interval. The four cross-fit and double-cross-fit estimates agree to about $0.006$ on the risk-difference scale at $r=100$. The full-sample AIPW reads lower ($0.033$, near g-computation). In the seed-1 diagnostic fit (Step~4), its in-sample residuals and one-step correction are smaller than out of fold. At $r=100$, DML's median is $0.043$, though without ground truth neither value can be called more accurate. Every interval excludes zero, reproducing the direction of the classic Connors et al. finding. The full-sample AIPW and TMLE intervals are reference Wald intervals; inference with the flexible nuisance library relies on the cross-fit rows (Step~5).}
\end{figure}

The punchline mostly holds, with one instructive exception. Figure~\ref{fig:rhc-panel} runs the same estimators at both a single cross-fit split (\(r=1\)) and the repeated-cross-fit median (\(r=100\)).
The four cross-fit and double-cross-fit routes agree tightly at \(r=100\): \(+0.043\) to \(+0.049\) on the risk-difference scale, a spread of about \(0.006\) (\(0.007\) at \(r=1\)).
Every interval excludes zero, and the direction reproduces the classic
\citet{connors1996} finding that right-heart catheterization is associated with increased
30-day mortality. On our risk-difference scale that is a four-to-five
percentage-point increase; \citet{connors1996} reported the association
as an odds ratio.

The exception is the \emph{full-sample} AIPW, which gives \(+0.033\), close to the g-computation estimate of \(+0.026\).
In the seed-1 residual diagnostic, the outcome model's residuals \(Y-\hat Q\) have root-mean-square \(0.30\) in sample and \(0.42\) out of fold.
The corresponding mean weighted corrections are \(+0.007\) and \(+0.014\), so the small in-sample correction leaves AIPW near the plug-in (companion code).
TMLE, which targets \(\hat Q\) instead of adding a correction, gives \(+0.053\) from the same initial in-sample fits.
We did not analyse its targeting step.

The seed-1 application run uses a different split from the one behind the \(r=1\) estimates of Figure~\ref{fig:rhc-panel}.
Its estimates are \(+0.044\) for cross-fitted AIPW (DML) and \(+0.047\) for CV-TMLE.
Their medians over 100 splits are \(+0.043\) and \(+0.047\), respectively.
The one-step and targeted estimates are thus closer out of fold, but without ground truth this does not establish which is more accurate.

On RHC, the fragile case is therefore the \emph{in-sample} fit, not the switch from a continuous to a binary outcome; the in-sample divergence is one reason to cross-fit flexible nuisances (Step~5).
Without ground truth, the out-of-fold agreement is an internal-consistency check, and the \citet{connors1996} finding whose direction it matches is itself subject to unmeasured confounding.

Two things change when the outcome is binary, and one thing does not.
What does not change is the estimand-solving logic: all three routes
still solve \(\mathbb{P}_{n} D = 0\) for the same efficient influence
function \(D\), and on RHC the cross-fit routes still agree. What changes is the machinery underneath.
First, the outcome nuisance \(Q(a,X)={P(Y=1\mid A=a,X)}\) is fit by a \emph{classification} Super Learner.
Second, TMLE's logistic fluctuation acts on the natural \(0/1\) scale of the outcome, with no rescaling, keeping the targeted \(Q\) in \((0,1)\).
The efficient influence function and its clever covariate keep the same form (Box~\ref{box:b10}).

\begin{biobox}[label={box:b10}]
For a binary outcome, the target is the risk difference
$\psi=P(Y^{1}=1)-P(Y^{0}=1)=E[Q_{1}(X)-Q_{0}(X)]$,
where $Q_{a}(X)=P(Y=1\mid A=a,X)$.
The efficient influence function keeps the same form as before, but now $Y\in\{0,1\}$ and $Q_{a}\in[0,1]$.

The outcome nuisance is therefore fit by a classification Super Learner.
Its base learners are trained for the binary outcome by classification-appropriate criteria.
The NNLS metalearner still combines them by squared error (Appendix~\ref{app:superlearner}).
The TMLE fluctuation is a genuine logistic submodel on the natural scale of the data.

Because TMLE is a substitution estimator evaluated at $\hat Q^{*}\in(0,1)$, its point estimate necessarily respects the $[-1,1]$ range of a risk difference.
The one-step AIPW point estimate carries no such guarantee.
The additive Wald interval, for either estimator, can still extend beyond $[-1,1]$.
That in-range property is one argument for a targeted plug-in when the outcome is rare or cells are sparse.
It does not by itself imply better inference \citep{qiu2024}.
RHC's common outcome (about $33\%$ 30-day mortality) does not represent the rare-outcome regime, and both point estimates lie well inside $[-1,1]$.
\end{biobox}

On RHC, the median-aggregated (\(r=100\)) double-cross-fit estimates, DC-AIPW \(+0.045\) and DC-TMLE \(+0.049\), lie within about \(0.002\) of their single-cross-fit counterparts
(DML \(+0.043\), CV-TMLE \(+0.047\); standard errors \(0.012\)--\(0.013\)
throughout). The medians agree, but a single double-cross-fit split varies more than a single cross-fit split (Step~5). Neither observation bears on interval coverage,
which no observational dataset without a known truth can establish.

The split-based rows appear at both \(r=1\) and \(r=100\) because a single arbitrary partition can move an estimate by more than the choice among these estimators does; Step~5 shows this on RHC and explains why we report prespecified medians over \(r=100\) splits.

\subsection{A Rosetta Stone for three
literatures}\label{a-rosetta-stone-for-three-literatures}

The literatures use different names, and sometimes different symbols, for the same objects, which is much of why the equivalence is easy to miss in practice.
Table~\ref{tab:rosetta} maps their notation, terminology, canonical references and software; Figures~\ref{fig:concept} and~\ref{fig:fitting-schemes} show the fitting schemes.

\begin{table}[!ht]
\centering
\caption{Rosetta Stone: three literatures, one set of objects. The AIPW/AIPTW, TMLE, and double-machine-learning traditions use different names, notation, references, and software for the same quantities.}
\label{tab:rosetta}
\begin{tabular}{@{}L{2.2cm}L{3.0cm}L{3.2cm}L{3.3cm}@{}}
\toprule
Object & AIPW / AIPTW & TMLE & Double machine learning \\
\midrule
Estimator & augmented IPW, one-step & targeted MLE (TMLE) & debiased / orthogonal-moment estimator \\
\addlinespace
What is solved & empirical mean of the EIF, added to the plug-in & $Q$ fluctuated until the EIF-mean is $0$ & orthogonal moment with out-of-fold nuisances \\
\addlinespace
Propensity $g(X)$ & propensity score & treatment mechanism $g$ & propensity, nuisance $m(X)$ \\
\addlinespace
Outcome $Q(a,X)$ & outcome regression & $\bar Q(a,X)$ & regression, nuisance $g(a,X)$ (DML's $g$ is the outcome model) \\
\addlinespace
Bias correction & augmentation term & clever-covariate fluctuation & Neyman-orthogonal score \\
\addlinespace
Canonical papers & \citet{robins1994}; \citet{bickel1993} & \citet{vanderlaan2006}; \citet{zheng2011} & \citet{chernozhukov2018}; \citet{newey2018} \\
\addlinespace
R packages & \texttt{AIPW} \citep{zhong2021} & \texttt{tmle} \citep{gruber2012}, \texttt{tmle3} \citep{coyle2021tmle3} & \texttt{DoubleML} \citep{bach2024} \\
\bottomrule
\end{tabular}
\end{table}

\textbf{Key takeaway.} The six doubly-robust estimators are one score on
a \(2\times3\) grid: two combination rules crossed with three fitting
schemes. On NHEFS and RHC, with the nuisances fit out of fold, they agreed to within finite-sample noise (in-sample fits can still diverge, as the RHC panel shows).

\begin{tcolorbox}[title={Step~4 in pseudocode --- three routes, one score}, colback=blue!3, colframe=blue!35, fonttitle=\bfseries, boxrule=0.4pt, arc=1pt]
AIPW: $\psi \leftarrow \mathrm{mean}_i[\hat Q_1 - \hat Q_0 + H\,(Y-\hat Q_A)]$; \quad $\mathrm{SE} \leftarrow \sqrt{\mathrm{var}(\hat D)/n}$ \\
TMLE: fit $\epsilon$ in $\mathrm{logit}\,\hat Q + \epsilon H$; \quad $\psi \leftarrow \mathrm{mean}(\hat Q^{*}_1 - \hat Q^{*}_0)$ \\
DML: the AIPW score, but $\hat g, \hat Q$ evaluated out-of-fold \quad \textit{\# all solve} $\mathbb{P}_n\hat D = 0$
\end{tcolorbox}

\section{Step~5 --- Why cross-fit, and a calibration
check}\label{step-5-why-cross-fit-and-a-calibration-check}

Step~4 showed AIPW, TMLE, and DML built from one efficient influence
function, and DML turned out to be nothing more than the AIPW score with
the propensity \(g(X)\) and outcome regression \(Q(a,X)\) fit \emph{out
of fold}. This raises the obvious applied question: why bother with the folds, when on NHEFS the point estimate barely moves? The answer is mainly about the confidence interval, although cross-fitting can also change point estimates materially (RHC, Step~4). We assess empirical interval coverage on a benchmark with known sample-average targets.

\subsection{Fitting and scoring on the same rows}\label{reusing-the-same-rows-to-fit-and-to-score-inflates-your-confidence}

With simple parametric nuisances you can fit \(g\) and \(Q\) on the
whole sample and evaluate the influence function on those same rows: a
Donsker condition then controls the empirical-process term. Flexible
machine-learning nuisances need not satisfy it. A random forest or Super
Learner can fit each observation partly to its own noise, so the
residual \(Y-\hat Q_A(X)\) scored at that observation comes from a model
that has already seen its \(Y\). This own-observation dependence can
bias the point estimate and distort its standard error
\citep{zivich2021, naimi2023}; the theory does not say in which
direction a given finite sample will move.

Cross-fitting removes the mechanism rather than patching it. Split the sample into \(V\) folds.
For each fold, fit \(\hat g\) and \(\hat Q\) on the \emph{other} folds and evaluate the score on the held-out fold.
No observation is used both to train a nuisance and to score its own residual.
Under the rate and positivity conditions in the box below, cross-fitting controls the empirical-process term without a Donsker condition \citep{chernozhukov2018} (Box~\ref{box:b11}).
That is why DML and CV-TMLE appear as separate rows from full-sample AIPW and TMLE: the same estimand and influence function, with nuisances fit out of fold.

Cross-fitting does not fix the finite-sample direction of bias, interval width, or coverage.
Full-sample fits can be valid under suitable conditions, and cross-fit intervals can undercover, cover near nominal, or be conservative.

\begin{biobox}[label={box:b11}]
\textbf{When is the analytic Wald interval valid?} Decompose the estimation error as in Box~\ref{box:b6},
\[
\hat\psi-\psi=\mathbb{P}_n D(O;g,Q,\psi)
+(\mathbb{P}_n-P)\{D(O;\hat g,\hat Q,\hat\psi)-D(O;g,Q,\psi)\}
+R_2 .
\]
The leading term satisfies $\sqrt{n}\,\mathbb{P}_n D\to N(0,\mathrm{Var}\,D)$.
The Wald interval $\hat\psi\pm1.96\,\widehat{\mathrm{SE}}$ is asymptotically valid when the other two terms are $o_P(n^{-1/2})$ and $n\,\widehat{\mathrm{SE}}^{2}\to\mathrm{Var}(D)$ in probability.

For independent observations and a fixed number of folds, the following conditions are sufficient:
\begin{enumerate}
\item[(i)] \textbf{Finite, positive EIF variance:} $0<\mathrm{Var}(D)<\infty$.
Finiteness holds, for example, for a bounded outcome under \emph{strong positivity}, $\varepsilon\le g\le 1-\varepsilon$.
\item[(ii)] \textbf{Bounded fitted propensities:} fitted propensities are bounded away from $0$ and $1$ (here by truncation).
\item[(iii)] \textbf{Consistency:} $\lVert\hat g-g\rVert$, $\lVert\hat Q-Q\rVert=o_P(1)$, so that the fitted score converges to $D$ in $L_2(P)$.
\item[(iv)] \textbf{Product rate:} $\lVert\hat g-g\rVert\,\lVert\hat Q_a-Q_a\rVert=o_P(n^{-1/2})$ for each arm $a$ (each $o_P(n^{-1/4})$ suffices).
This condition bounds $R_2$ (Box~\ref{box:b6}).
\item[(v)] \textbf{Empirical-process control:} the empirical-process term is controlled by \emph{cross-fitting} or by a Donsker condition on the nuisance class.
\end{enumerate}

\textbf{Targeted estimators.}
For TMLE these conditions apply to the targeted fit $\hat Q^{*}$, whose empirical score must also be $o_P(n^{-1/2})$ (Box~\ref{box:b8}).
For CV-TMLE and pooled DC-TMLE, cross-fitting applies to the initial nuisance fits.
Because the pooled fluctuation uses held-out outcomes, condition (v) additionally requires stochastic equicontinuity of the one-dimensional fluctuation class at the targeted fit.
Together with targeted-score $L_2(P)$ consistency, this controls the remaining empirical-process term.

\textbf{Full-sample rows.}
The full-sample AIPW/TMLE rows use flexible learners (\texttt{ranger}, \texttt{earth}), so condition~(v) would have to come from a Donsker condition.
The in-sample empirical-process term need not vanish for such learners, because $\hat g,\hat Q$ depend on the very points where the score is evaluated (own-observation overfitting).
Cross-fitting breaks that dependence and removes the need for the Donsker condition.
We therefore read the full-sample rows as reference baselines and rely on the cross-fit rows for inference.
\end{biobox}

\subsection{A calibration check where the truth is
known}\label{a-calibration-check-where-the-truth-is-known}

NHEFS has no ground truth, so it cannot tell us whether a 95\% interval covers 95\% of the time.
For that we use the IHDP (Infant Health and Development Program) semi-synthetic benchmark \citep{hill2011}.
It generates outcomes from known response surfaces on real covariates and has adequate overlap for comparing the estimators (companion code).
Its design differs from the sampling model of Steps~3--5 in two respects.

First, all 50 realizations share one fixed set of \(n=747\) covariate profiles and treatment assignments.
Only the simulated response surfaces and outcomes change between realizations.
Second, the target of realization \(b\) is the sample average of its response surfaces \(\mu_{1b}\) and \(\mu_{0b}\),
\[
\psi_b=\frac1n\sum_{i=1}^n\{\mu_{1b}(X_i)-\mu_{0b}(X_i)\},
\]
which ranges from \(3.16\) to \(12.65\) across realizations.
The frequency with which an interval includes \(\psi_b\) is therefore coverage conditional on this design, for a sample-average target.
We report this inclusion frequency as coverage.

This fixed-design experiment does not directly assess unconditional coverage for a fixed superpopulation ATE, the target for which the Wald interval of Box~\ref{box:b11} is built.
Such an assessment would require a sampling design or averaging argument corresponding to that target.

The superpopulation EIF variance includes variation in the conditional treatment effect across covariate profiles, although its average is fixed under this design \citep[Sections~4.1--4.2]{balzer2016}.
\citet{balzer2016} derive this decomposition for trials in which the covariates are held fixed but treatment remains randomized.
Here the treatment assignments are fixed as well, so we borrow only the decomposition.
With oracle nuisances and a conditional central limit theorem, this component makes the usual EIF interval conservative for the sample-average target when effects vary across profiles.
This mechanism may contribute to high inclusion frequencies, but does not by itself explain the observed coverage of any fitted route.

\begin{figure}[htbp]\centering
\includegraphics[width=1.00\linewidth]{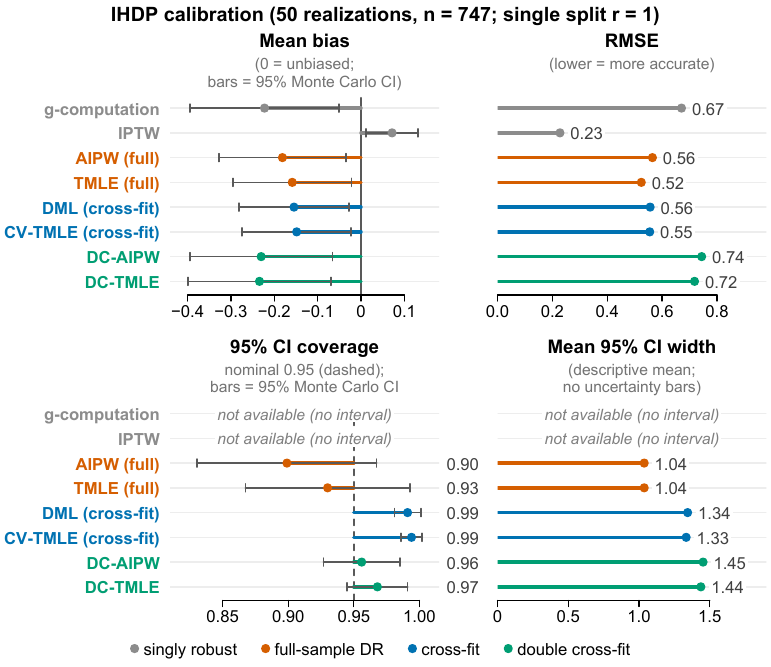}
\caption{\label{fig:ihdp-grading}IHDP calibration of all eight estimators on 50 outcome realizations, generated independently conditional on one fixed design of covariates and treatment assignments ($n=747$). Panels show mean bias, RMSE, nominal 95\% interval coverage of each realization's known sample-average target, and mean 95\% interval width; colours distinguish nuisance-fitting strategies. Split-based results average 20 fold partitions per realization, each scored separately at a single split ($r=1$); they are not results for a median-aggregated estimator. Full-sample methods were also refitted 20 times per realization. Their performance measures are averaged over these runs. Bias and coverage bars are 95\% Monte Carlo confidence intervals with the realization as the independent unit \citep{morris2019}; RMSE and width have no uncertainty bars. G-computation and IPTW have no reported analytic interval, so their coverage and width are unavailable. Full-sample inclusion frequencies are not detectably below 0.95 (the AIPW shortfall is $\approx$1.5 MCSE). Step~5 gives the paired comparisons and explains why this experiment does not directly assess unconditional coverage for the superpopulation ATE.}
\end{figure}

Given the design, the outcome realizations are generated independently.
The realization-level Monte Carlo standard errors used below therefore apply to this conditional coverage.
Running all eight estimators on the 50 realizations, each under 20 random fold partitions, and grading every interval against its \(\psi_b\) gives the calibration summary of Figure~\ref{fig:ihdp-grading}: mean bias, RMSE, coverage and mean interval width (the companion code).
Three counts must be kept apart.

\textbf{Outcome realizations:} the 50 realizations are the distinct simulated outcome sets.
They are the Monte Carlo replicates from which bias, RMSE (root-mean-square error), coverage and mean interval width are computed.

\textbf{Fold partitions:} the 20 partitions are independent random fold assignments \emph{within} one realization.
They are the same kind of object as a cross-fitting repetition, but used differently.
Here each partition is scored on its own as a single split (\(r=1\)).
Averaging the 20 coverage indicators reduces Monte Carlo variation from the fold assignments when estimating single-split coverage.

\textbf{Repeated-split aggregation:} scoring the partitions separately is \emph{not} repeated cross-fitting, which would instead \emph{combine} them into one estimate and one interval, a different estimator (next subsection).
In short, the 20 partitions \emph{measure} the \(r=1\) estimator; repetition would \emph{change} it.

The figure compares the routes, and the Monte Carlo intervals show the precision of the bias and coverage estimates.
Paired comparisons below assess differences between routes.

\textbf{Coverage differs by fitting scheme.}
At \(n=747\), the cross-fit DML and CV-TMLE intervals cover \(0.99\).
Their mean widths are \(1.34\) and \(1.33\), about 30\% wider than the full-sample intervals (mean width \(1.04\)).
Full-sample AIPW and TMLE cover \(0.90\) and \(0.93\).
DC-AIPW and DC-TMLE cover \(0.96\) and \(0.97\), with still wider intervals (\(1.45\) and \(1.44\)), so width alone does not order coverage.

The right Monte Carlo unit is the realization, not the fold partition.
We treat the 50 realizations as independent replicates given the design, with coverage averaged over the 20 partitions within each.
The paired AIPW\(\to\)DML coverage gap is \(0.09\), with \(t=2.8\) across realizations.
The TMLE\(\to\)CV-TMLE gap is \(0.06\), with \(t\approx2.0\).
The first gap is resolved; the second is borderline.

DC-AIPW covers \(0.035\) less often than DML (paired \(t\approx-2.9\)).
DC-TMLE covers \(0.026\) less often than CV-TMLE (paired \(t\approx-2.5\)).
Their gaps above the corresponding full-sample estimators are not resolved (\(t\approx1.7\) and \(1.2\)).

The full-sample inclusion frequencies are not detectably below 0.95.
Full-sample AIPW, at 0.90, sits only about \(1.5\) Monte Carlo standard errors below it.
This comparison does not assess coverage for the superpopulation ATE.
So the data do not show cross-fitting repairing a demonstrated under-coverage.
They show out-of-fold fitting raising coverage on this design, resolved for AIPW\(\to\)DML.
This benchmark does not explain why the cross-fit routes reach \(0.99\).

\textbf{On bias and RMSE, the differences are small or borderline.}
Mean bias is $-0.18$ for full-sample AIPW, $-0.16$ for TMLE, $-0.15$ for both cross-fit routes, and $-0.23$ for both double-cross-fit routes.
Each has a Monte Carlo standard error of 0.06--0.08, so their marginal intervals in Figure~\ref{fig:ihdp-grading} all overlap.

Paired over realizations, two small bias differences are resolved.
TMLE is less biased than AIPW by 0.02 (paired $t\approx3.2$), and CV-TMLE is less biased than DML by 0.006 ($t\approx2.8$).
No other bias difference among the doubly-robust routes is resolved ($|t|<1.9$).

RMSE is 0.52--0.56 for the full-sample and cross-fit routes and 0.72--0.74 for the double-cross-fit routes.
The paired MSE difference between each double-cross-fit route and its full-sample counterpart sits about two Monte Carlo standard errors from zero: $t\approx2.0$ for DC-AIPW versus AIPW and $t\approx2.1$ for DC-TMLE versus TMLE.
Against the cross-fit routes it sits less far ($t\approx1.8$ versus DML and $1.6$ versus CV-TMLE).
The singly-robust g-computation baseline has bias $-0.22$ and RMSE 0.67, and is shown without an analytic interval.

At one data-generating process, one sample size, and 50 realizations, these are calibration observations, not a leaderboard.

The double-cross-fit rows need one design note. DC-AIPW and DC-TMLE
fit \(g\) and \(Q\) on \emph{disjoint} thirds of the data, so at \(n=747\) each nuisance is trained on \(249\) units, against about \(598\) for the five-fold cross-fit routes. We do not attribute the larger RMSE or width of the double-cross-fit rows to this smaller training sample, or to their different number of folds: the benchmark changes both at once and was not designed to separate them.

On the larger RHC cohort of Step~4 (Figure~\ref{fig:rhc-panel}), the median-aggregated double-cross-fit estimates agree with the other split-based routes, while their single splits vary more (Step~5).
RHC has no ground truth, so it says nothing about their coverage.

Bias and coverage are accompanied by realization-level Monte Carlo standard errors (Box~\ref{box:b12}), and differences between routes are assessed with paired, realization-level comparisons.
The first 10 realizations give an AIPW\(\to\)DML coverage gap of \(0.275\) (paired MCSE \(0.123\)), compared with about \(0.09\) over all 50.
This illustrates how unstable a small pilot can be.
With 50 realizations, the paired AIPW\(\to\)DML coverage gap is resolved.
Among the doubly-robust routes, the two small paired bias differences reported above are resolved; the other bias differences are not.
The reported paired MSE contrasts are about two Monte Carlo standard errors or fewer from zero.

\begin{biobox}[label={box:b12}, break at=-\baselineskip/0pt]
Coverage here is a two-level Monte Carlo quantity.
The $B=50$ outcome realizations are generated independently on one fixed design, and each is analysed under $S=20$ fold partitions of the \emph{same} dataset.
Let $(L_{bs},U_{bs})$ be the interval from realization $b$ and partition $s$, and let
\[
\bar C_b=\frac1S\sum_s\mathbf 1\{L_{bs}\le\psi_b\le U_{bs}\}
\]
be realization $b$'s coverage averaged over partitions.

The 20 partitions are not independent replicates, so a naive binomial $\sqrt{\hat\pi(1-\hat\pi)/(BS)}$, which treats all $BS$ intervals as independent, understates the uncertainty.
A binomial $B=50$ formula errs the other way.
It treats each realization as a single pass or fail, ignores the averaging over partitions and overstates the Monte Carlo standard error.
Treat the realization as the unit instead, with the realization-level mean and standard error:
$$
\hat\pi=\frac{1}{B}\sum_{b=1}^{B}\bar C_b, \qquad \mathrm{MCSE}(\hat\pi)=\frac{\mathrm{sd}(\bar C_1,\dots,\bar C_B)}{\sqrt{B}} .
$$

This realization-clustered MCSE is $0.032$--$0.035$ for the full-sample fits, whose coverages are $0.93$ and $0.90$.
It is $0.012$--$0.015$ for double cross-fitting and approximately $0.004$--$0.005$ for single cross-fitting.
Averaged over the 50 realizations, coverage is $0.991$ for DML and $0.994$ for CV-TMLE.

Differences between routes are compared within realization, as paired differences over the 50 realizations (Step~5).
Any reported coverage should travel with its design ($B$, $S$) and its clustered MCSE.
\end{biobox}

\subsection{Repeating the split: how many repetitions are
enough?}\label{repeating-the-split-how-many-repetitions-are-enough}

The IHDP grading of Figure~\ref{fig:ihdp-grading} scored every cross-fit and double-cross-fit estimate on a \emph{single} random partition of the data (\(r=1\)). Because the fold
assignment is arbitrary, that one split injects a small, avoidable
source of variation into the point estimate.
The hazard is not only accidental.
An analyst can re-draw seeds until an estimate crosses a threshold (a \(p\)-value, a sign, a narrower interval), so reporting a single hand-picked split is a subtle and hard-to-detect form of \(p\)-hacking \citep{zivich2024seed, schader2024seed, naimi2024seed}.
\citet{karim2025splits} and \citet{zivich2021} recommend \emph{repeating} the split \(r\) times with independent seeds and median-aggregating, since the median is robust to an outlying partition where the mean is not.
The point estimate becomes the median of the \(r\) per-split estimates \(\hat\psi_s\).
Each split's variance \(\hat V_s\) (of the estimator itself) is augmented by its squared distance from that median before the median is taken:
\[
\tilde\psi_r=\operatorname{median}_{s\le r}\hat\psi_s,\qquad
\tilde V_r=\operatorname{median}_{s\le r}\bigl\{\hat V_s+(\hat\psi_s-\tilde\psi_r)^{2}\bigr\},\qquad
\tilde\psi_r\pm1.96\sqrt{\tilde V_r}.
\]
Our code uses this estimator-scale aggregation, following \citet{zivich2021}. As an exploratory check of this aggregated estimator on IHDP, we combined each realization's 20 partitions by the rule (\(r=20\)). For each of the four split-based routes, the aggregated interval included \(\psi_b\) in all 50 realizations (95\% Clopper--Pearson interval \(0.93\)--\(1.00\)). Fifty pass-or-fail outcomes cannot grade its coverage more finely, and the design caveat above applies. The practical
questions are how large \(r\) must be, and where it matters.

\begin{figure}[htbp]\centering
\includegraphics[width=1.00\linewidth]{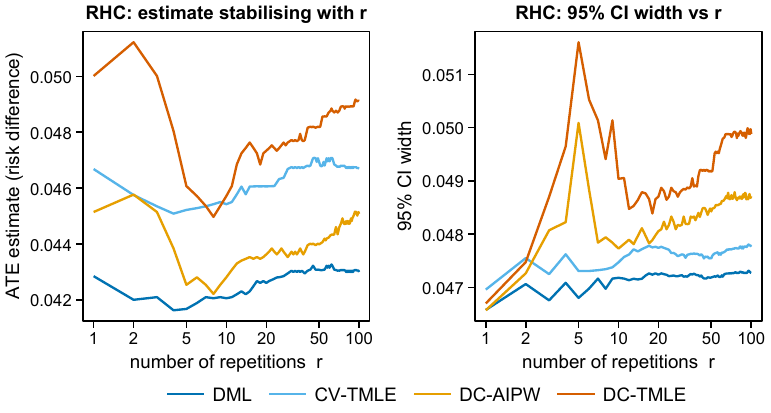}
\caption{\label{fig:rhc-reps}RHC repetition study ($n=5735$; $100$ repeated cross-fits, median-aggregated). Left: the median-aggregated ATE estimate (risk difference) versus the number of repetitions $r$. The single-cross-fit medians (DML, CV-TMLE) stay within $0.001$ of their $r=100$ value from $r=7$ and $r=12$ on. The double-cross-fit medians (DC-AIPW, DC-TMLE) do so only from $r=54$ and $r=41$, after excursions of up to $0.004$. From $r=1$ to $r=100$, no estimate moves by more than $0.0009$ net. Right: the $95\%$ CI width grows by $1.5$--$6.8\%$ from $r=1$ to $r=100$; RHC has no known truth, so coverage is not assessed.}
\end{figure}

We examine them on the real, large-$n$ RHC data of Step~4, where double cross-fitting is actually deployed \citep{mondol2025}.
We rerun the four split-based estimators with up to $r=100$ repeated cross-fits, reporting the running median at every intermediate $r$ (Figure~\ref{fig:rhc-reps}).

\textbf{Variation across individual splits.}
Across the 100 splits, the single-split estimates have standard deviations of 0.0017 for DML and 0.0021 for CV-TMLE on the risk-difference scale (companion code).
Double-cross-fit splits scatter more: 0.0030 for DC-AIPW and 0.0037 for DC-TMLE.
The single-split DML estimate ranges from 0.038 to 0.047, while its median is a stable 0.043.
That swing, about 0.0095, exceeds the entire 0.006 spread among the four split-based estimators at \(r=100\) (Step~4), so a single split can move an estimate by more than the choice among these estimators does.
A reader comparing one arbitrary split against another could therefore mistake partition noise for an ``estimator difference.''

\textbf{Variation in running medians.}
The double-cross-fit medians wander further and settle later.
Their largest distance from the $r=100$ value is 0.0029 for DC-AIPW and 0.0041 for DC-TMLE, against 0.0014 for DML and 0.0016 for CV-TMLE.
DML and CV-TMLE stay within 0.001 of their $r=100$ values from $r=7$ and $r=12$ onward.
DC-AIPW and DC-TMLE do so only from $r=54$ and $r=41$.

In this sequence, the first split happened to land close to the final median for all four estimators.
The largest net change from $r=1$ to $r=100$ is 0.0009, for DC-TMLE.
By contrast, at $r=8$ the double-cross-fit medians sat up to 0.004 below their final values.
The standard deviations above show how far a different first split could have landed.

\textbf{Interval widths.}
From $r=1$ to $r=100$, widths grow by 1.5\% for DML, 1.7\% for CV-TMLE, 4.6\% for DC-AIPW, and 6.8\% for DC-TMLE (Figure~\ref{fig:rhc-reps}, right).
RHC has no known truth, so this does not assess the coverage of the aggregated intervals.

The guidance we draw is correspondingly narrow, because it rests on one dataset and one sequence of random splits.
Repeat the split and median-aggregate for single and double cross-fitting alike, with \(r\) fixed in advance, as in the Step~4 panels (\(r=100\)).
Do so especially for double cross-fitting, whose single splits scattered most here.
On these data, \(r=100\) was ample.

How many repetitions another problem needs is a question one dataset cannot settle, so we defer it to the simulation study of \citet{karim2025splits}.
There, for double-cross-fit TMLE at \(n=3000\), \(r=25\) performed as well as \(r=100\), whereas \(r=1\) differed visibly.
Across cross-fit and double-cross-fit TMLE and AIPW, \citet{karimfu2026} found that repetition reduced dependence on the partition but made several intervals more conservative.
\citet{ellul2025} applied the same median rule over 50 splits.
With their reduced learner library, aggregation lowered coverage below that of a single split.
Repetition aims to reduce sensitivity to the partition.
It does not guarantee improved coverage.

\textbf{Key takeaway.} With machine-learning nuisances, build the Wald interval from cross-fit nuisance fits; Box~\ref{box:b11} gives the conditions under which it is valid. Repeat the split and median-aggregate to reduce split sensitivity, especially for double cross-fitting. Report any coverage with its replicate count and Monte Carlo standard error.

\begin{tcolorbox}[title=Step~5 in pseudocode --- cross-fit and calibrate, colback=blue!3, colframe=blue!35, fonttitle=\bfseries, boxrule=0.4pt, arc=1pt]
\texttt{for} fold $k$ \texttt{in} $1..V$: fit $\hat g, \hat Q$ on the other folds, score $\hat D$ on fold $k$ \quad \textit{\# cross-fitting} \\
$\Rightarrow$ no own-observation overfit; valid SE for machine-learning nuisances under the conditions of Box~\ref{box:b11} \\
grade on IHDP (fixed design, known sample-average target), 50 realizations: bias, RMSE, coverage $\pm$ MCSE
\end{tcolorbox}

\section{Step~6 --- Sensitivity and failure modes
(positivity)}\label{step-6-sensitivity-and-failure-modes-positivity}

The NHEFS results illustrate close agreement under good practical overlap.
We now contrast NHEFS with the poor-overlap LaLonde/NSW--PSID example, and then examine the sensitivity of the NHEFS estimate to analysis choices.

\subsection{Poor overlap in LaLonde/NSW--PSID}\label{where-everything-breaks-the-lalondenswpsid-positivity-failure}

The clearest way to see the failure is to move to a dataset known to
lack overlap. The LaLonde/NSW--PSID construction \citep{lalonde1986}
compares job-training participants (the National Supported Work
demonstration, NSW) against a general-population comparison group (the
Panel Study of Income Dynamics, PSID) whose covariates barely resemble
the participants'. Our purpose is to show how sensitive the estimate becomes when overlap is this poor: to the estimator and its fitting scheme, to the random seed, and to the overlap region retained by trimming. Two things are worth stating up front.

The estimand here is the pooled average treatment effect over this PSID-dominated sample.
The well-known \$1,794 figure is a different quantity: the training effect in the \citet{dehejia1999} subsample of LaLonde's \emph{experimental} NSW sample, a well-overlapping population.
Because it concerns a different population, we do not use it to judge the sign, bias, or coverage of the pooled-sample estimates.

\begin{figure}[htbp]\centering
\includegraphics[width=\linewidth]{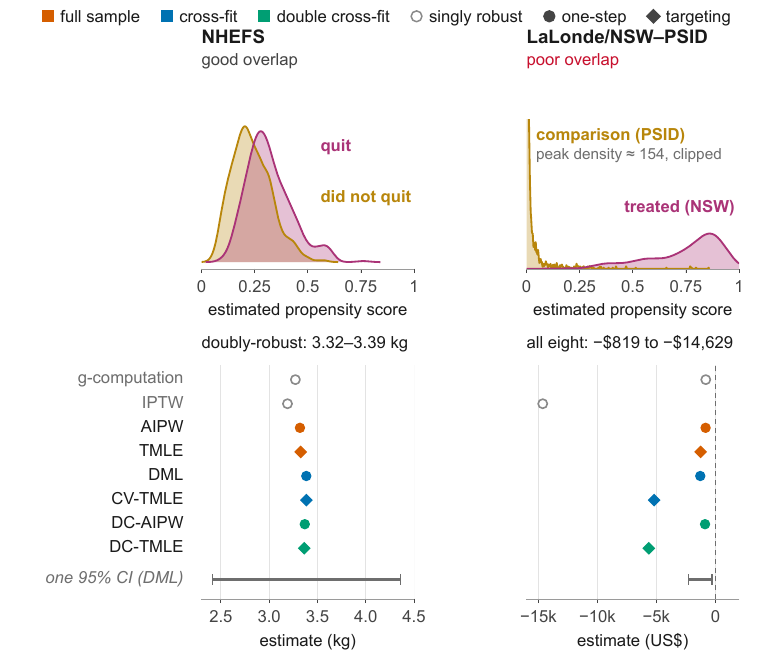}
\caption{\label{fig:positivity}Overlap and estimates in NHEFS (left) and LaLonde/NSW--PSID (right). Upper panels show kernel densities of estimated propensities from separate three-learner diagnostic Super Learners (logistic regression, lasso and random forest), not the analysis fits. NHEFS uses the archived diagnostic of Figure~\ref{fig:nhefs-overlap}; LaLonde uses a seeded diagnostic fit to the original data (companion code). The LaLonde comparison-group density peaks near zero at about 154 and is clipped for display. Lower panels show all eight point estimates on each dataset's outcome scale: 1971--1982 weight change in kg for NHEFS and 1978 earnings in US dollars for LaLonde. For split-based estimators, NHEFS uses medians over 100 partitions and LaLonde uses one partition (Table~\ref{tab:positivity-full}). The grey bars show one nominal 95\% DML interval per dataset, aggregated over the 100 partitions for NHEFS and from one partition for LaLonde. Neither population has a known effect. This descriptive comparison does not rank accuracy or isolate overlap from other differences between the datasets.}
\end{figure}

The NSW--PSID design also makes a \emph{structural} positivity violation plausible.
NSW admitted only disadvantaged applicants who met its eligibility criteria \citep{lalonde1986}, so many PSID comparison units could not have qualified for the program. The estimated propensities cannot confirm this on their own (Step~1). What they do show is practical overlap so poor that, whatever its cause, the pooled ATE cannot be estimated stably from these data without extrapolation.
Figure~\ref{fig:positivity} sets this overlap, and the spread of the eight estimates, beside those of NHEFS. On the right, the treated and comparison
densities are nearly disjoint.

Tables~\ref{tab:positivity-full} and~\ref{tab:positivity} show what the doubly-robust machinery does with that. On the full, untrimmed sample the cross-fit AIPW score (which is exactly DML) returns \(-\$1{,}284\) with a 95\% interval of \((-\$2{,}278,\ -\$291)\) that lies entirely below zero.
This is not an AIPW-specific artefact, and sample-splitting is no remedy.
All six doubly-robust estimators in Table~\ref{tab:positivity-full} use propensity-based corrections in this poor-overlap setting, and all return negative estimates, whether the initial nuisances are fitted on the full sample, cross-fitted or double-cross-fitted.
The propensity bound is active here: it clips 2{,}093 of 2{,}675 predictions in the audited double-cross-fit run.
Truncation limits the weights but does not resolve a structural positivity violation.
Out of fold, the \emph{TMLE} estimates move furthest from the full-sample ones, to \(-\$5{,}186\) (CV-TMLE) and \(-\$5{,}644\) (DC-TMLE).
We did not run a diagnostic that separates the roles of weak support, the initial fits, and the targeting step in these shifts.

A seed sweep confirms that the exact figures shift but the sign, and the position far below the full-sample estimates, do not.
Across eight seeds the cross-fit CV-TMLE has a median of \(-\$5{,}077\) (range \(-\$3{,}641\) to \(-\$5{,}384\)), and DC-TMLE a median of \(-\$4{,}830\) (range \(-\$4{,}125\) to \(-\$5{,}644\)). What buys
valid inference under good overlap (Step~5) is no protection against poor overlap.

\begin{table}[!ht]
\centering
\caption{Poor overlap on LaLonde/NSW--PSID affects every route. Each estimator on the full, untrimmed sample; the \$1{,}794 experimental estimate concerns a different population and is not a benchmark for these estimates. Split-based rows use one random split. The ordering here is not a ranking: every route returns a negative estimate, and with no known truth for this population, no estimate can be called more or less accurate than another.}
\label{tab:positivity-full}
\begin{tabular}{llr}
\toprule
Estimator & Nuisance fitting & $\hat\psi$ (US\$) \\
\midrule
g-computation & full-sample      & $-819$ \\
IPTW          & full-sample      & $-14{,}629$ \\
AIPW          & full-sample      & $-833$ \\
TMLE          & full-sample      & $-1{,}254$ \\
AIPW $=$ DML  & cross-fit        & $-1{,}284$ \\
CV-TMLE       & cross-fit        & $-5{,}186$ \\
DC-AIPW       & double-cross-fit & $-881$ \\
DC-TMLE       & double-cross-fit & $-5{,}644$ \\
\bottomrule
\end{tabular}
\end{table}

Restricting to the region where the two groups' estimated propensities
overlap changes both the population and the answer (Table
\ref{tab:positivity}). Trimming to \(\hat g \in [0.05, 0.95]\) keeps 423 of 2,675 units (16\%) and gives \(+\$484\).
Trimming to \([0.10, 0.90]\) keeps 300 units (11\%) and gives \(-\$27\).
Their interval half-widths are roughly \$1,700--\$2,200, and both intervals include zero.
These shifts show sensitivity to a drastic change in the retained population, bought by discarding 84--89\% of the sample \citep{petersen2012}.
They neither recover nor validate the pooled ATE. You cannot
manufacture overlap the data do not contain. Trimming reports this by
shrinking the population to the sliver of estimated common support,
not by recovering the effect for everyone.

\begin{table}[!ht]
\centering
\caption{Sensitivity of LaLonde/NSW--PSID to estimated-propensity trimming. Each row refits the cross-fit AIPW score ($=$ DML) after applying the stated rule, so the retained population changes. Intervals are nominal, computed within the selected sample, and do not account for estimating the selection rule.}
\label{tab:positivity}
\begin{tabular}{l r r r c}
\toprule
Trimming rule & $n$ kept & \% kept & $\hat\psi$ (US\$) & 95\% CI (US\$) \\
\midrule
None (all data)          & 2{,}675 & 100 & $-1{,}284$ & $(-2{,}278,\ -291)$ \\
$\hat g \in [0.05, 0.95]$ & 423     & 16  & $+484$     & $(-1{,}229,\ 2{,}196)$ \\
$\hat g \in [0.10, 0.90]$ & 300     & 11  & $-27$      & $(-2{,}214,\ 2{,}161)$ \\
\bottomrule
\end{tabular}
\end{table}

\begin{biobox}[label={box:b13}]
Weak overlap inflates both the efficiency bound and the second-order remainder of every route (Box~\ref{box:b3}). If the bound is infinite, no regular asymptotically linear estimator of $\psi$ with finite asymptotic variance exists, whatever the route (AIPW, TMLE, or DML). As Box~\ref{box:b3} shows, propensities approaching $0$ or $1$ do not by themselves make it infinite. The remainder's multiplier, of order $1/\{\hat g(X)(1-\hat g(X))\}$, rests on propensities that the sparse data near the boundary cannot pin down.

Trimming to $\{x : \lambda \le g(x) \le 1-\lambda\}$ does not repair this; it changes the target from $\psi$ to a trimmed estimand
$$ \psi_\lambda = E\!\left[\, Q_1(X) - Q_0(X) \;\middle|\; \lambda \le g(X) \le 1-\lambda \,\right], $$
the effect in a different population \citep{petersen2012}. This oracle target is defined by the true propensity. Our illustration instead selects units by estimated propensities and refits the nuisances on the retained sample.
Treating that selection as fixed gives a descriptive sensitivity interval, not inference for $\psi_\lambda$.
Inference for $\psi_\lambda$ would also have to account for estimating the boundary. Either way the question has narrowed to the 11--16\% of units the trimming retains; it has not recovered $\psi$. 
\end{biobox}

\subsection{Two sensitivity checks on NHEFS (good
overlap)}\label{two-sensitivity-checks-on-nhefs-good-overlap}

Back on NHEFS the picture is the opposite.
Within our engine, the six doubly-robust estimates span 0.07 kg when the split-based estimates are median-aggregated over \(r=100\) (Table~\ref{tab:knobs}).
Two further checks, both drawn from the companion code, concern the truncation bound and the software implementation.

\textbf{Propensity-truncation threshold.} Our engine truncates the
estimated propensity to \([0.025, 0.975]\) before forming weights.
No propensities were clipped at these bounds in the audited analysis fits (Step~1).
Loosening the bounds therefore leaves those fitted estimates unchanged.
Exercise~1 (Appendix~\ref{app:exercises}) asks the reader to assess tighter bounds and to contrast truncation with LaLonde's population-changing trimming.
The LaLonde case differs in kind: trimming there discards observations rather than capping weights, and the propensity range it keeps changes the answer entirely (Table~\ref{tab:positivity}).

\textbf{Software implementation.} The second check runs the same estimand through four independent packages alongside our Super Learner engine \citep{vanderlaan2007, gruber2012, zhong2021, bach2024, coyle2021tmle3}. Run once each, the five implementations keep the NHEFS estimate within 3.32--3.49 kg, a 0.18 kg envelope (Table~\ref{tab:knobs}, last row). The 0.18 kg spread is about one fifth of the \(\approx 0.9\) kg confidence half-width. Each point estimate also lies within that half-width of the published 3.4--3.5 kg benchmark \citep{hernan2020}. The envelope shows robustness, not attribution; Step~7 examines what can account for the spread.

\textbf{Key takeaway.} On NHEFS (good overlap) the answer is robust to estimator and software implementation, and the truncation bound clips no propensity in the audited fits. Under poor overlap (LaLonde) no route estimates the target stably without extrapolation; cross-fitting does not rescue the failure, and trimming relabels the estimand rather than recovering it.

\begin{tcolorbox}[title=Step~6 in pseudocode --- sensitivity and failure, colback=blue!3, colframe=blue!35, fonttitle=\bfseries, boxrule=0.4pt, arc=1pt]
vary the truncation window and the software $\rightarrow$ re-estimate $\psi$ \quad \mbox{\textit{\# sensitivity}} \\
on LaLonde (poor overlap): trim to $\hat g \in [\lambda, 1-\lambda]$ $\rightarrow$ discards 84--89\% of units \\
\textit{\# trimming changes the estimand; it does not recover} $\psi$
\end{tcolorbox}

\section{Step~7 --- Reporting checklist and software
reconciliation}\label{step-7-reporting-checklist-and-software-reconciliation}

Everything up to here has argued a single point: AIPW/AIPTW, TMLE, and
DML are three routes to the same efficient influence function for the
ATE, and on NHEFS and RHC, with the nuisances fit out of fold, they returned the same answer up to finite-sample noise (Step~4). That is reassuring, but it leaves a working analyst
with two practical questions. First, \emph{what do I actually write in
the Methods section} so a reader can reproduce and audit the estimate?
Second, \emph{my colleague ran the \textup{\texttt{tmle}} package and I ran \textup{\texttt{DoubleML}}, and we got different kilograms --- who is right?}
This step answers both. It is the use-value of the tutorial: a one-page
reporting checklist, and a software reconciliation that turns an
apparent contradiction between packages into a short list of settings to document and investigate. 

\subsection{A reporting checklist for doubly-robust and double-ML
analyses}\label{a-reporting-checklist-for-doubly-robust-and-double-ml-analyses}

The organizing thesis of this tutorial --- \emph{diagnose overlap first}
--- is really a claim about reporting. The decisions that determine
whether a doubly-robust estimate is trustworthy are made \emph{before}
the point estimate exists, and most of them are invisible unless the
analyst writes them down. Table \ref{tab:checklist} is that list, filled
in for the NHEFS analysis. Reading it top to bottom recovers the whole
pipeline: the estimand and its three identification assumptions; the
Super Learner library \citep{vanderlaan2007} shared by both nuisances;
whether and how the nuisances were cross-fit; the propensity truncation
and its rationale \citep{petersen2012}; the overlap diagnostic that was
inspected before estimation; how sensitive the estimate is to trimming;
the estimator finally reported and why; and where the standard error
came from.

Two rows carry most of the auditing weight. The \emph{truncation} row makes the analyst commit to a bound (here \(\hat g(X)\in[0.025,0.975]\)) and state its purpose, so a reader can see whether a few near-zero denominators were allowed to inflate the influence-function variance.
The \emph{SE-source} row forces the distinction between a Wald interval read off the influence function and a bootstrap interval.
For the doubly-robust and double-ML estimators the former suffices under the conditions of Box~\ref{box:b11}.
This is why the panel reports analytic intervals for them and withholds intervals from the singly-robust baselines. The checklist's numbers come from the companion code and its output files.

\begin{table}[!htp]
\centering
\setstretch{1}%
\caption{Reporting checklist for a doubly-robust / double-machine-learning ATE analysis, filled in for the NHEFS example. Every row is a choice a reader must be able to see to reproduce or audit the estimate.}
\label{tab:checklist}
\begin{tabular}{@{}L{0.235\linewidth}L{0.735\linewidth}@{}}
\toprule
Reporting item & Specification for the NHEFS analysis (and rationale) \\
\midrule
Estimand & ATE $\psi = E[Y^{1}-Y^{0}]$: effect of smoking cessation ($A$) on 1971--1982 weight change in kg ($Y$); $n=1566$, 403 quitters. \\
Identification assumptions & (i) consistency; (ii) conditional exchangeability $Y^{a}\indep A\mid X$ (no unmeasured confounding); (iii) positivity $0<g(X)<1$. \\
Super Learner library & \texttt{SL.mean}, \texttt{SL.glm}, \texttt{SL.glmnet} (lasso), \texttt{SL.ranger} (random forest), and \texttt{SL.earth} (MARS). The same library fits both the propensity $g(X)$ and the outcome regression $Q(a,X)$ \citep{vanderlaan2007}. \\
Cross-fitting (folds, repeated splits) & Full-sample fits (classic AIPW/TMLE); single cross-fit, $V=5$ (DML/CV-TMLE); double cross-fit, $V=3$, with $g$ and $Q$ trained on disjoint folds (DC-AIPW/DC-TMLE) \citep{chernozhukov2018, zivich2021}. We report $r=1$ estimates and the median over $r=100$ seeded repetitions (Step~5). Single-split estimates scatter with an SD of about 0.08~kg for the cross-fit and 0.14~kg for the double-cross-fit routes (Step~4). \\
Propensity truncation & $\hat g(X)$ is bounded to $[0.025,\,0.975]$, which bounds the clever covariate, $|\hat H|\le40$ \citep{petersen2012}. This caps each inverse-probability weight but does not guarantee balanced relative influence (Box~\ref{box:b8}). \\
Overlap diagnostic & Propensity-score density by treatment group (Figure~\ref{fig:nhefs-overlap}), inspected \emph{before} estimation. This is the ``diagnose overlap first'' step. \\
Trimmed fraction and sensitivity & Good overlap: in the audited fits (Step~1), the $[0.025,0.975]$ bounds bind on none of the 1566 observations, so truncation leaves those estimates unchanged. The 0.07~kg span of the six doubly-robust estimates at $r=100$ is variation across estimators and fitting schemes, not truncation sensitivity. By contrast, in the positivity-failure vignette (Step~6), trimming to common support can discard 84--89\% of the sample. \\
Chosen estimator and rationale & Primary: CV-TMLE (single cross-fit TMLE). It is an in-range plug-in, with cross-fit inference valid for ML nuisances under the conditions of Box~\ref{box:b11}. AIPW/DML and the double-cross-fit forms share its EIF and are reported alongside it; all six lie within 0.07~kg at $r=100$. \\
Standard-error source & EIF-analytic Wald, with no bootstrap: $\widehat{\mathrm{SE}}=\sqrt{\mathbb{P}_n\{\hat D^{2}\}/n}$ from the influence function. DML uses out-of-fold fits; CV-TMLE uses out-of-fold initial fits followed by pooled targeting (Box~\ref{box:b15}). For estimates aggregated over $r$ splits, the median rule of Step~5 applies. The g-computation and IPTW point estimates are shown without intervals. These flexible plug-ins have no generally justified analytic interval here, and a naive bootstrap would not remove their possible first-order bias (Box~\ref{box:b15}). \\
\bottomrule
\end{tabular}
\end{table}

\begin{biobox}[label={box:b15}]
The analytic standard error is read off the estimated efficient influence function $\hat D$, so no resampling is required:
\[
\widehat{\mathrm{SE}}(\hat\psi)=\sqrt{\mathbb{P}_n\{\hat D^{2}\}/n}.
\]
The Wald interval is $\hat\psi\pm1.96\,\widehat{\mathrm{SE}}(\hat\psi)$.
For an estimate aggregated over repeated splits, the interval uses the median rule of Step~5 instead.

Which fits enter $\hat D$ depends on the estimator.
For AIPW and DML, $\hat D$ is the same estimated EIF that the one-step estimator uses; for DML it is built from out-of-fold nuisance fits.
For TMLE, $\hat D$ uses the targeted $\hat Q^{*}$, which is exactly the fit for which $\mathbb{P}_n\{\hat H(A,X)(Y-\hat Q^{*}_A)\}=0$.
For CV-TMLE and pooled DC-TMLE, it uses out-of-fold initial fits followed by pooled targeting, and their validity is governed by the targeted-score conditions of Box~\ref{box:b11}.

The singly-robust g-computation and IPTW estimators have no comparable orthogonal influence-function variance.
For parametric fits, model-based or sandwich intervals can be derived under the corresponding conditions.
For the flexible plug-ins used here, no generally justified off-the-shelf root-$n$ interval exists, and a naive bootstrap does not remove their possible first-order bias (Box~\ref{box:b1}).
Their intervals are therefore omitted from the panel.
\end{biobox}

\subsection{Software reconciliation: one estimand, five
implementations}\label{software-reconciliation-one-estimand-five-implementations}

We put the identical NHEFS estimand through five independent implementations: the hand-coded engine of this paper, packaged as the \texttt{dr3} R package; the \texttt{tmle} package \citep{gruber2010, gruber2012}; the \texttt{AIPW} package \citep{zhong2021}; \texttt{DoubleML} \citep{bach2024}; and \texttt{tmle3} from the tlverse \citep{coyle2021tmle3}.
Table~\ref{tab:software} collects the results (the companion code).
The estimates span 3.32 to 3.49~kg, a range of 0.18~kg computed before rounding.
All sit in the 3.3--3.5~kg region consistent with the canonical Hernán--Robins NHEFS benchmark \citep{hernan2020}.
The entries differ in several settings at once, so the table alone cannot say which setting moves the number.

The learner library is largely shared.
We gave the \texttt{tmle} and \texttt{AIPW} packages our five-learner Super Learner library, and \texttt{tmle3} an \texttt{sl3} stack of the same five learners.
\texttt{DoubleML} used a random forest for both nuisances.
Truncation is inactive in the audited NHEFS fits.
The estimated propensities of the overlap diagnostic lie well inside the $[0.025,0.975]$ bound (Table~\ref{tab:nhefs-overlap}), as do those of the analysis fits we audited (Step~1).

The fitting scheme sorts the table into two groups.
Our full-sample AIPW and TMLE fit and predict both nuisances in-sample and give 3.32--3.33~kg.
Every other entry uses out-of-fold outcome predictions and gives 3.38--3.49~kg.
Our DML, the \texttt{AIPW} package, and \texttt{DoubleML} use five-fold cross-fitting.
At its default, the \texttt{tmle} package (version 2.1.1) uses cross-validated initial outcome predictions with a full-sample propensity.
The \texttt{tmle3} entry uses ten-fold CV-TMLE.

Our own engine shows the same grouping (Step~4): at $r=100$, every split-based estimate (3.36--3.39~kg) lies above the full-sample ones.

Finally, each entry is a single run, so it carries single-split noise.
That noise comprises one draw of the cross-fitting split, where there is one, and of every other random choice (internal cross-validation folds, random forests).

\begin{table}[!ht]
\centering
\caption{The same NHEFS estimand (smoking cessation on weight change, kg) across five independent implementations, one row per reported estimate. Each entry is a single run. The columns distinguish the combination rule from nuisance fitting, including hybrid schemes outside the grid of Figure~\ref{fig:concept}. A one-step estimate with both nuisances cross-fitted is DML in this paper's terminology. Estimates span 3.32--3.49 kg (a 0.18 kg range); the text finds the spread compatible with the fitting scheme plus single-split noise, without appeal to the estimator label or truncation. The hand-coded cross-fit (DML) entry is one cross-fit run (seed 1) for package comparability; Figure~\ref{fig:nhefs-panel} plots a different single split, and the median over $r=100$ repeated cross-fits is $3.38$ kg. The hand-coded DML, \texttt{tmle}, and \texttt{AIPW} entries are the first of the 20 seeded runs in Figure~\ref{fig:software-seeds}. The hand-coded engine of this paper is packaged as the \texttt{dr3} R package.}
\label{tab:software}
\begin{tabular}{@{}L{0.26\linewidth}L{0.15\linewidth}L{0.33\linewidth}r@{}}
\toprule
Implementation & Combination rule & Nuisance-fitting scheme & $\hat\psi$ (kg) \\
\midrule
Hand-coded (\texttt{dr3}) & One-step & Full sample & 3.32 \\
Hand-coded (\texttt{dr3}) & Targeting & Full sample & 3.33 \\
Hand-coded (\texttt{dr3}) & One-step & Five-fold cross-fit & 3.41 \\
\texttt{tmle} \citep{gruber2012} & Targeting & Out-of-fold initial $\hat Q$; full-sample $\hat g$ & 3.38 \\
\texttt{AIPW} \citep{zhong2021} & One-step & Five-fold cross-fit & 3.49 \\
\texttt{DoubleML} \citep{bach2024} & One-step & Five-fold cross-fit & 3.46 \\
\texttt{tmle3} \citep{coyle2021tmle3} & Targeting & Cross-fit initial fits (CV-TMLE) & 3.40 \\
\bottomrule
\end{tabular}
\end{table}

\begin{figure}[htbp]\centering
\includegraphics[width=\linewidth]{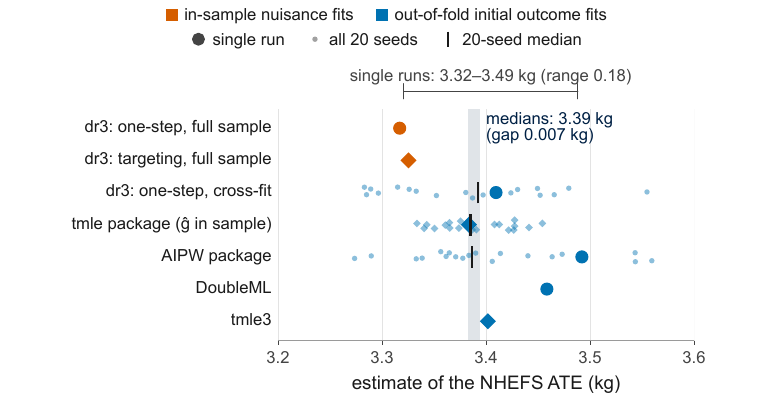}
\caption{\label{fig:software-seeds}Single runs and seed-to-seed variation of the NHEFS estimate (smoking cessation on 1971--1982 weight change, kg; $n=1566$) across five implementations. Large symbols are the single runs of Table~\ref{tab:software}; \texttt{dr3} is this paper's hand-coded engine. Small symbols are all 20 seeded runs, the first included, of the three implementations built on our \texttt{SuperLearner} library, each with an explicit 0.025 propensity bound. Black ticks mark the medians, which all round to 3.39~kg and differ by at most 0.007~kg. The pale band highlights their location. Circles are one-step and diamonds targeted estimates. Colour separates in-sample from out-of-fold initial outcome fits; the \texttt{tmle} package fits its propensity model on the full sample. \texttt{DoubleML} (random forests) and \texttt{tmle3} were run once. The comparison describes variation on these data and does not isolate the effect of any single setting.}
\end{figure}

To size that noise, we re-ran the three implementations built on our Super Learner library (our DML, the \texttt{AIPW} package, and the \texttt{tmle} package) with 20 seeds each and an explicit 0.025 propensity bound (the companion code).
The first of these runs reproduces their Table~\ref{tab:software} entries exactly.
That includes the \texttt{tmle} entry, which had been computed under the package's own default bound, so the bound does not bind there either.
Figure~\ref{fig:software-seeds} plots all 20 runs of each beside the single runs of Table~\ref{tab:software}.

From one seed to the next, a single five-fold cross-fit moves by about 0.08~kg, its SD across seeds.
The seeds redraw the folds, the internal cross-validation, and the random forests.
The \texttt{AIPW} package alone ranges from 3.27 to 3.56~kg.
The \texttt{tmle} package varies less, with SD about 0.04~kg.

The medians nonetheless agree to within 0.01~kg (largest gap 0.007~kg): our DML, the \texttt{AIPW} package, and the \texttt{tmle} package's TMLE all give 3.39~kg.
The 0.18~kg spread of the single runs is about two of the cross-fit SDs.
Both of its ends are AIPW estimates.
The lower end is our full-sample fit (3.32~kg), about 0.07~kg below the cross-fit medians.
The upper end is the \texttt{AIPW} package's first seed (3.49~kg), about 0.11~kg above its own median.

The spread is therefore compatible with the fitting scheme plus single-split noise.
Neither the estimator label nor truncation is needed to account for it.

The reconciliation has limits.
A shared learner library does not produce identical nuisance predictions.
Packages draw different internal cross-validation folds and estimate different ensemble weights.
The \texttt{tmle} package also uses its own fitting scheme.
TMLE updates $\hat Q$ by targeting, whereas AIPW adds a one-step correction.

Together, these differences yield medians within 0.01~kg of one another on NHEFS.
Their individual effects were not isolated and could offset one another.
We ran \texttt{DoubleML} and \texttt{tmle3} only once.
Their estimates, 3.46 and 3.40~kg, fall inside the seed-to-seed range above.
They are consistent with this explanation but do not test it.
In particular, this comparison does not isolate \texttt{DoubleML}'s learner-library effect.

This is the reconciliation the practitioner needs. When a \texttt{tmle}
run (3.38 kg) and a \texttt{DoubleML} run (3.46 kg) disagree, the gap, about 0.075~kg before rounding, is not evidence that TMLE and DML target different quantities: under the conditions of Box~\ref{box:b11} they share one influence function and are asymptotically equivalent \citep{kennedy2023, hines2022, diaz2020}. Nor can one pair of runs say which setting produced the gap. The two runs differ in estimator, learner library, and fitting scheme at once, and a gap of this size is within single-split noise. A workable order has three steps.
First, compare fitting schemes \citep{zheng2011, chernozhukov2018}.
Then compare medians over repeated splits rather than single runs \citep{karim2025splits}.
Then investigate any remaining gap by changing one implementation choice at a time while holding the other inputs fixed.

The checklist and the software table are two views of one discipline.
The checklist says which settings to disclose.
The software table shows why those settings are what a reader needs.
The software spread is compatible with differences in fitting scheme and single-split variation, both recorded in the cross-fitting row. Disclosing them is a precondition for reproducing a doubly-robust or double-ML estimate across packages, and that cross-package reproducibility is the practical contribution of this tutorial.

Appendix~\ref{app:exercises} provides five exercises using the companion code, covering truncation, learner-library ablation, coverage, repeated splitting and package reconciliation.

\textbf{Key takeaway.} Disclose the learner library, the fitting scheme and folds, the number of repeated splits, and the truncation bound, and compare packages on medians over repeated splits rather than single runs.

\begin{tcolorbox}[title=Step~7 in pseudocode --- report and reconcile, colback=blue!3, colframe=blue!35, fonttitle=\bfseries, boxrule=0.4pt, arc=1pt]
report: estimand, three assumptions, SL library, folds, repeated splits, truncation, overlap diagnostic, SE source \\
run the same estimand through \{\texttt{tmle}, \texttt{AIPW}, \texttt{DoubleML}, \texttt{tmle3}\} \\
\textit{\# compare medians over repeated seeds, not single runs}
\end{tcolorbox}

\section{Discussion}\label{discussion}

Set side by side on a shared nuisance library and matched initial fits within each fitting scheme, AIPW, TMLE, and DML do not behave like rival estimators competing for the right answer; the dominant pattern is agreement.
On NHEFS, the six doubly-robust estimates span \(3.32\) to \(3.39\)~kg at \(r=100\), a range of \(0.07\)~kg (Figure~\ref{fig:nhefs-panel}).
Every doubly-robust interval covers the canonical Hernán--Robins benchmark of roughly \(3.4\) to \(3.5\)~kg \citep{hernan2020}.

The same convergence repeats on the second real dataset, right-heart catheterization (Figure~\ref{fig:rhc-panel}), once the nuisances are fit out of fold.
There the cross-fit routes agree on a small positive mortality effect on the risk-difference scale.
Every route reproduces the direction of \citet{connors1996}.
On the IHDP known-truth benchmark, coverage separates by fitting scheme (below), but the routes stay close in bias.
Mean biases run from \(-0.15\) to \(-0.23\), and the only resolved paired differences are at most \(0.023\).

AIPW, TMLE, and DML are three routes to the same efficient influence function (EIF).
Under the conditions of Box~\ref{box:b11}, they are asymptotically equivalent and efficient \citep{kennedy2023, hines2022, diaz2020, robins1994, bickel1993}.
The examples illustrate finite-sample agreement under the established asymptotic theory.
In the displayed NHEFS and RHC panels, matched out-of-fold one-step and targeted estimates differ by about one tenth of a confidence-interval width or less.
Across individual RHC partitions, the gap can reach about 0.15 of the mean interval width of the pair.
Differences are larger for the in-sample RHC fits and the poor-overlap LaLonde example (Steps~4 and~6).

\subsection{What actually matters: diagnose overlap
first}\label{what-actually-matters-diagnose-overlap-first}

If the estimator label is not the decision that matters most, the diagnostic
that precedes it is. Step~1 of our workflow is to look at the
estimated-propensity distributions by treatment group before fitting any
effect. The estimated-propensity plot shows substantial overlap in NHEFS (Figure~\ref{fig:nhefs-overlap}).
The agreement in Figure~\ref{fig:nhefs-panel} rests on that overlap (a precondition, not a guarantee; Step~4) and is no substitute for checking it.

The counter-case makes the point sharper.
In the LaLonde/PSID positivity vignette the treated and control propensity distributions barely intersect (Figure~\ref{fig:positivity}).
The full-sample estimators then span \(-\$819\) (g-computation) to \(-\$14{,}629\) (IPTW), all negative and differing by more than an order of magnitude. Restricting to the estimated common-support region discards \(84\) to \(89\%\) of the sample and leaves intervals roughly \(\pm\$2{,}000\) wide, which include effects of either sign and carry almost no information
(Table~\ref{tab:positivity}).
Trimming does not recover the answer; it reports that outside the region of overlap the effect cannot be estimated without extrapolation \citep{petersen2012}. No amount of
doubly-robust machinery, flexible learning, or cross-fitting
manufactures information that the design did not collect.

\subsection{Cross-fitting and inference}\label{cross-fitting-buys-valid-inference-not-a-different-estimand}

The second lever that matters is cross-fitting, examined in Step~5.
Its role is inferential: under the rate and positivity conditions of Step~5 (Box~\ref{box:b11}), which it does not itself supply, it yields valid standard errors when the nuisances are flexible machine-learning fits, without changing the estimand.
Its cost is extra variance across fold assignments.
Its benefit is that flexible nuisances outside a Donsker class, whose estimation errors converge more slowly than \(n^{-1/2}\), still yield valid \(\sqrt n\) inference when their errors meet the product-rate condition of Step~5.

On the IHDP benchmark (Step~5, Figure~\ref{fig:ihdp-grading}), the full-sample inclusion frequencies were not detectably below 0.95.
The higher coverage of the cross-fit routes is therefore best read not as cross-fitting repairing a demonstrated under-coverage, but as out-of-fold estimation raising coverage on this design.
These are coverages conditional on one fixed design, for a sample-average target, not superpopulation coverage.

Double cross-fitting is a further refinement: for each held-out fold, it fits the initial propensity and outcome regressions on two disjoint training folds \citep{newey2018, zivich2021, mondol2025}.
On NHEFS the double-cross-fit estimates (\(3.37\) and \(3.36\)~kg at \(r=100\)) sit inside the same \(0.07\)-kg band as the rest (Figure~\ref{fig:nhefs-panel}).
Their single splits, however, scatter more (Step~4).

\subsection{Reconciling the software: same estimand, different settings}\label{reconciling-the-software-same-estimand-different-defaults}

The five implementations of Table~\ref{tab:software} print estimates that are close but not identical.
On NHEFS their spread is compatible with the fitting scheme and single-split noise, without appeal to the estimator label or truncation (Step~7).
A reader who gets \(3.49\)~kg from one package and \(3.38\)~kg from another should therefore check the fitting scheme and compare medians over repeated splits before reading the gap as a substantive disagreement.

\subsection{A short guide: which route
when}\label{a-short-guide-which-route-when}

Because the routes share one EIF, choosing among them is about
finite-sample behaviour and diagnostics, not about a different target.
Table~\ref{tab:whichwhen} distills the guidance; it is a
lean-toward table, not a ranking, and every row assumes overlap has
already been checked.

\begin{table}[!ht]
\centering
\caption{A short guide to leaning toward a doubly-robust route. All routes share one efficient influence function; the choice concerns finite-sample behaviour and diagnostics, not the estimand.}
\label{tab:whichwhen}
\begin{tabular}{@{}L{0.44\linewidth}L{0.52\linewidth}@{}}
\toprule
Situation & Lean toward \\
\midrule
Strong overlap, low-dimensional parametric nuisance models & AIPW (one-step; analytic EIF standard error) \\
Rare outcome or sparse cells, want in-range estimates & TMLE / CV-TMLE (bounded plug-in) \\
Flexible ML nuisances, moderate-to-large $n$ & DML or CV-TMLE (cross-fitting for valid inference; repeat the split) \citep{chernozhukov2018, zivich2021} \\
High-dimensional confounding, $g$--$Q$ dependence a concern & DC-AIPW / DC-TMLE (double cross-fit, if a third of the data trains each nuisance adequately; repeat the split) \citep{newey2018, mondol2025, karim2025splits} \\
Near-positivity violation & none blindly --- diagnose overlap, truncate, report sensitivity \citep{petersen2012} \\
\bottomrule
\end{tabular}
\end{table}

\subsection{Relation to prior work}\label{relation-to-prior-work}

This tutorial complements and extends several strands of the applied
literature. It is a hands-on counterpart to the code-free comparative
review of \citet{moccia2024}, supplying the end-to-end implementation
that a code-free treatment omits. It generalizes \citet{mondol2025},
which worked a single estimator (DC-TMLE) on the RHC data, to the
three-route, one-EIF view across two real datasets. It sits alongside
the targeted-learning and doubly-robust tutorials of \citet{schuler2017}
and \citet{luque2018}, the cross-fitting exposition of
\citet{zivich2021}, and recent guidance on machine learning for causal
inference in epidemiology \citep{naimi2023, renson2025}. It is closest
in spirit to recent tutorials that treat
these tools one at a time: the reproducible multi-language walk-through
of \citet{smith2022}, the TMLE tutorials of \citet{luque2018} and
\citet{talbot2025} (the latter for time-to-event outcomes), and the
method-selection guide of \citet{keele2025}. This tutorial differs in three ways.
First, it unifies AIPW, TMLE, and DML as three routes to a single influence function.
Second, it runs all three routes, including their cross-fit versions, through one shared nuisance library, with matched initial fits within each fitting scheme.
Third, it grades them against a known-truth benchmark.

Closest in subject matter are four simulation-based comparison studies.

\citet{tan2022} compared AIPW, TMLE, double-score matching, and the penalized-spline method PENCOMP, with regression imputation and IPTW as comparators.
Each was fit with generalized linear, generalized additive, or Super Learner nuisance models.
Their analyses included simulations, a plasmode study, and a fibromyalgia cohort.
Overall, AIPW and TMLE with Super Learner had the smallest bias and MSE.
With generalized linear or additive nuisance models, TMLE was the more robust of the two in one of their four synthetic scenarios.
In the plasmode study, the two were similar.
The authors recommend TMLE with Super Learner nuisances and bootstrap standard errors.
Their ``double machine learning'' means any doubly-robust estimator with machine-learning nuisances, not the cross-fitted construction of \citet{chernozhukov2018} that we call DML.
None of the estimators they describe splits the sample.

\citet{ellul2025} compared AIPW and TMLE with Super Learner in simulations based on a birth cohort with high-dimensional confounding ($n$ from 200 to 2000; 14 or 87 confounders), with and without cross-fitting.
The two performed similarly in most settings, but TMLE was more stable: cross-fitted AIPW gave implausible results on a few datasets.
Cross-fitting mattered more for standard-error estimation and coverage than for point estimates.
The number of folds mattered less than whether to cross-fit at all.
The more flexible of their two Super Learner libraries reduced bias in their most complex scenarios.
Without cross-fitting, however, it worsened standard-error estimation and coverage.
A sensitivity analysis aggregated estimates over 50 repeated splits by the median rule of Step~5, in their most complex scenario and in the case study.
It left their conclusions about the number of folds unchanged.
With their reduced (less flexible) library, aggregation gave more biased model-based standard errors and lower coverage than a single split, at $n=200$ and at $n=2000$.
The aggregated case-study intervals were wider.

Both studies suggest, without testing it, that TMLE's advantage reflects its being a substitution estimator whose estimates stay within the parameter's range.
\citet{luque2018} had already noted this finite-sample difference.

\citet{meng2026refine2} offer REFINE2, a tool that simulates plasmode data from an analyst's own covariates.
It grades IPW, AIPW, TMLE, and double-cross-fit AIPW and TMLE by bias and coverage.
In their case study, singly-fit AIPW and TMLE with non-smooth learners undercovered badly.
Double cross-fitting brought coverage near or above nominal in two of their three scenarios but was not always beneficial.

\citet{karimfu2026} compare TMLE and AIPW constructions while holding the learner library, inner folds, and truncation bound fixed.
The constructions differ in cross-fitting, targeting, and repetition.

These studies ask which estimator and which nuisance-fitting choices perform best in simulation, and none compares the major software packages for these estimators side by side. This tutorial asks a different question: when and why the routes agree, and what separates the numbers that different implementations print.
Against that backdrop, the specific practitioner novelty here is the software-bridge reconciliation of Table~\ref{tab:software}.
It carries one estimand through five implementations.
The three built on the same \texttt{SuperLearner} library are re-run over 20 seeds to size their single-split noise (Step~7).

\subsection{Extensions: time-to-event
outcomes}\label{extensions-time-to-event-outcomes}

The three-routes logic is not specific to a continuous or binary
outcome. For a time-to-event outcome the target is typically a survival
or risk difference at a fixed horizon, or a difference in restricted
mean survival time. Each such estimand has its own efficient influence function, with a censoring model among the nuisances and additional censoring and positivity conditions.
One-step, targeted, and cross-fit estimators exist for these targets.
A causal hazard ratio needs its own careful definition, and the ATE score derived here does not apply to it unchanged.
The workflow carries over: diagnose overlap first, then any doubly-robust route, with cross-fitting for flexible learners under conditions analogous to those of Step~5. Readers who need the
survival case in full should consult the step-by-step
TMLE-and-machine-learning tutorial of \citet{talbot2025} and the
continuous-time targeted estimators of \citet{rytgaard2023}.
Software includes the discrete-time \texttt{survtmle} \citep{benkeser_survtmle} and the
continuous-time \texttt{concrete} \citep{chen_concrete} packages.

\subsection{Further reading}\label{further-reading}

For readers who want to go deeper: \citet{hernan2020} for the
causal-inference workflow and the g-formula; \citet{vanderlaan2007} for
the Super Learner; \citet{vanderlaan2006} for TMLE; \citet{kennedy2023} for a
modern semiparametric review of the doubly-robust /
efficient-influence-function machinery; and \citet{chernozhukov2018} for
double machine learning and cross-fitting. For worked, step-by-step
tutorials of the individual routes, see \citet{luque2018} and
\citet{smith2022}, and, for time-to-event outcomes, \citet{talbot2025}.

\subsection{Limitations}\label{limitations}

Three limits should temper how far these results are read.

\textbf{Benchmark scope.}
The known-truth evidence is pedagogical rather than a claim of external validity.
IHDP is a semi-synthetic benchmark with an investigator-specified response surface \citep{hill2011}.
Its coverage numbers speak to estimator behaviour under a controlled data-generating process, not to the transportability of any particular effect.
They are also conditional on one fixed design and a sample-average target (Step~5).

\textbf{Real-data scope.}
The main walkthrough uses two real datasets in parallel: NHEFS for a continuous outcome (weight change) and RHC for a binary outcome (30-day mortality).
LaLonde/NSW--PSID provides a separate poor-overlap illustration.
These examples cover a limited range of designs rather than a broad survey, and the qualitative patterns could differ under other confounding structures.

\textbf{Precision and repetition.}
The finite-sample bias contrasts come from a single generating process at a single sample size and should not be generalized into a ranking.
On IHDP, cross-fit DML shows a small mean bias ($-0.15$, MCSE $\approx 0.065$).
Full-sample AIPW shows a larger one at $-0.18$ (MCSE $\approx 0.075$), and the double-cross-fit routes are at $-0.23$.
The DML--AIPW gap is a fraction of one MCSE and is not resolved even when paired over realizations ($t\approx1.2$), so the ordering is suggestive at most (Figure~\ref{fig:ihdp-grading}).

All four split-based routes in this grading (DML, CV-TMLE, DC-AIPW, DC-TMLE) were moreover scored as single splits ($r=1$), not as the repeated, median-aggregated estimators we recommend (Steps~4--5).
Those were checked only in an exploratory way (Step~5).
The repetition evidence of Step~5 (Figure~\ref{fig:rhc-reps}) is similarly narrow: one dataset and one sequence of random splits.
The equivalence we lean on is asymptotic, and the finite-sample departures we observed are the positivity- and cross-fitting-dependent effects catalogued above.

\subsection{Conclusion}\label{conclusion}

Three routes, one answer: AIPW, TMLE, and DML solve the same efficient-influence-function equation for the ATE.
Cross-fitted AIPW and DML that use the same score and inputs are the same estimator; TMLE uses a different combination rule.
The worked examples show why the estimator label alone is insufficient to interpret software differences.
On NHEFS, the six doubly-robust estimates span \(0.07\)~kg when the split-based estimates are median-aggregated over 100 splits (Figure~\ref{fig:nhefs-panel}).
The three implementations sharing our \texttt{SuperLearner} library have medians within \(0.01\)~kg over 20 seeds (Figure~\ref{fig:software-seeds}).
In-sample RHC fits and the poor-overlap LaLonde example show the limits of that agreement.
Agreement does not establish accuracy, and the IHDP coverage evidence remains conditional on one fixed design.

The choices that carry real weight come before and around the estimator.
Diagnose overlap before choosing an estimator (Step~1).
Cross-fit flexible nuisances and prespecify the number of repeated splits; valid inference still requires the rate and regularity conditions of Box~\ref{box:b11}.
Use double cross-fitting only when a third of the data is an adequate training sample for each nuisance, and assess its split sensitivity by repeating the split.
Report the estimand, learner library, fitting scheme, truncation, targeting and aggregation rules, and standard-error calculation.
Accompany simulation-based bias and coverage comparisons with Monte Carlo standard errors.
For applied researchers, the operative message is the reframing from ``which estimator'' to ``do the data support the estimate, and is the inference valid''.

\section*{Acknowledgements}
This research was supported in part through computational resources and services provided by Advanced Research Computing at the University of British Columbia. During this work the author used generative AI tools, including Anthropic's Claude, to assist with writing analysis and simulation code, and revising text. The author reviewed and verified all outputs and takes full responsibility for the content.

\section*{Author's statements}

\subsection*{Funding information}
This research received no specific grant from any funding agency in the public, commercial, or not-for-profit sectors.

\subsection*{Conflict of interest}
The author declares no competing interests.

\subsection*{Ethical approval and informed consent}
Not applicable. Only publicly available, de-identified secondary data and a semi-synthetic benchmark were analysed; no participants were recruited and no new data were collected. Under Article~2.2 of the Tri-Council Policy Statement: Ethical Conduct for Research Involving Humans (TCPS~2), research that relies exclusively on publicly available information does not require research ethics board review.

\subsection*{Consent for publication}
Not applicable.

\subsection*{Data availability statement}
All four datasets analysed are publicly available: NHEFS \citep{hernan2020}, right-heart catheterization \citep{connors1996}, LaLonde/NSW--PSID \citep{lalonde1986,dehejia1999} and the IHDP semi-synthetic benchmark \citep{hill2011}.
They are not redistributed with this article or its companion code.
The companion repository's README and website give each dataset's source.
For the datasets read from files, they also give the expected file names and their MD5 checksums.

\subsection*{Code availability}
Code and reproduction materials are available as follows.
\begin{itemize}
\item Code implementing the estimators, the software reconciliation, and every figure and table, with the scripts that ran the replicate studies and the archived result files they produced:\\ \url{https://github.com/ehsanx/dml-alternatives}.
\item A browsable reproduction from the archived result files, with a numeric audit of the reported numbers:\\ \url{https://ehsanx.github.io/dml-alternatives/}.
\item The companion R package \texttt{dr3}, which wraps the estimators in a single interface:\\ \url{https://github.com/ehsanx/dr3}; documentation and vignettes:\\ \url{https://ehsanx.github.io/dr3/}.
\end{itemize}
Code regenerates the reported tables, figures and numerical summaries from the archived results; the analysis scripts are also provided.
Refitting the stochastic analyses from the raw data need not reproduce the archived values exactly, because Super Learner fits depend on learner-package versions and on the computing platform.
The repository's README and website record the computing environments and the fits that were not seeded.

\subsection*{Author contribution}
M.E.K. conceived the study, wrote the software, performed the analyses, and wrote and revised the manuscript.

\clearpage
\appendix
\section{Super Learner weight selection and the oracle guarantee}\label{app:superlearner}

This appendix states how the Super Learner of Step~2 chooses its ensemble weights, and what its oracle guarantee does and does not cover.
Box~\ref{box:b4} summarizes the consequences for the estimators.

\textbf{Ensemble and cross-validated risk.}
Write the Super Learner for $Q$ (identically for $g$) as a convex combination of the $J$ library learners $\hat Q^{(1)},\dots,\hat Q^{(J)}$,
$$
\hat Q_{\alpha} \;=\; \sum_{j=1}^{J}\alpha_j\,\hat Q^{(j)},
\qquad
\alpha\in\Delta^{J-1}=\Big\{\alpha:\ \alpha_j\ge 0,\ \textstyle\sum_{j}\alpha_j=1\Big\}.
$$
Let $\mathcal V_k$ be the $k$th of the $V$ validation folds, and let $\hat Q^{(j)}_{-k}$ be the $j$th learner trained with fold $k$ held out.
For a loss $L$ (specified below), the $V$-fold cross-validated risk of a weight vector $\beta$ is
$$
\widehat R_{\mathrm{CV}}(\beta) \;=\;
\sum_{k=1}^{V}\sum_{i\in\mathcal V_k}
L\!\left(Y_i,\ \sum_{j}\beta_j\,\hat Q^{(j)}_{-k}(A_i,X_i)\right).
$$

\textbf{Weight fitting.}
The simplex selector minimizes the cross-validated risk over the whole simplex, that is, over non-negative weights that sum to one:
\[
\hat\alpha^{\mathrm{CC}}
=\arg\min_{\alpha\in\Delta^{J-1}}\widehat R_{\mathrm{CV}}(\alpha).
\]
The package calls this \texttt{method.CC\_LS}.

Our engine uses the \texttt{SuperLearner} default \texttt{method.NNLS} instead.
It first computes, without the sum-to-one constraint,
\[
\hat\beta=\arg\min_{\beta\ge0}\widehat R_{\mathrm{CV}}(\beta).
\]
If $\sum_j\hat\beta_j>0$, it normalizes these coefficients to obtain
$\hat\alpha^{\mathrm{NNLS}}=\hat\beta/\sum_j\hat\beta_j$.
If every $\hat\beta_j$ is zero, the package warns that all algorithms have zero weight and keeps the zero weights.
Normalized NNLS is a common approximation to $\hat\alpha^{\mathrm{CC}}$, but it does not in general minimize $\widehat R_{\mathrm{CV}}$ over the simplex.

\textbf{Prediction loss.}
In both selectors, $L$ is the squared-error loss for $Q$ and $g$.
Declaring $g$ binomial changes how the base learners are fitted.
It does not change the metalearner's squared-error criterion that combines them.

\textbf{Oracle guarantee.}
The oracle inequality concerns selection among finitely many candidate weight vectors.
An example is a grid on the simplex whose number of points grows at most polynomially with $n$.
The excess prediction risk of a combination is its risk for predicting new outcomes minus that of the true $Q$, with the candidates fitted on the training folds.
The oracle is the candidate with the smallest true risk; the selected combination is the candidate with the smallest cross-validated risk.
Under a bounded loss, the expected excess prediction risk of the selected combination is within a factor arbitrarily close to one of the oracle's expected excess risk, plus a remainder of order $\log n/n$ \citep{vanderlaan2007}.
If the oracle's expected excess risk $r_n$ satisfies $nr_n/\log n\to\infty$, the selected combination and the oracle have asymptotically equivalent expected excess risk.

As the grid is refined, grid selectors approach $\hat\alpha^{\mathrm{CC}}$ in cross-validated risk.
The result therefore bears on $\hat\alpha^{\mathrm{CC}}$ through that approximation, and on normalized NNLS not directly.
Good prediction risk does not by itself deliver the nuisance rates that inference on $\psi$ needs (Box~\ref{box:b11}).

\section{Exercises}\label{app:exercises}

Each exercise re-uses the shared nuisance engine and data of the
accompanying repository. Notes in parentheses give expected findings where our analyses support them, and otherwise say what to compare.

\begin{enumerate}
\def\labelenumi{\arabic{enumi}.}
\tightlist
\item
  \textbf{Truncation sensitivity.} Recompute the NHEFS panel with the
  propensity-truncation window tightened to \([0.05,0.95]\) and loosened
  to \([0.01,0.99]\). Hold the untruncated initial nuisance predictions and the fold assignments fixed while varying the bound. Recompute each estimator from them, including the targeting step where applicable, so that stochastic refitting is not confused with truncation sensitivity. How much does \(\hat\psi\) move? Then contrast this truncation, which caps weights and keeps every observation, with the LaLonde/PSID trimming of Step~6, which discards observations. (Compare the numbers of clipped propensities and the resulting estimates. The LaLonde trimming changes the retained population and with it the estimand.)
\item
  \textbf{Library ablation.} Drop \texttt{SL.ranger} and
  \texttt{SL.earth}, leaving \texttt{SL.mean}, \texttt{SL.glm},
  \texttt{SL.glmnet}. Does the doubly-robust estimate change by more
  than the estimator-to-estimator spread of
  Figure~\ref{fig:nhefs-panel}? (Our runs do not settle this: none of them isolates the library (Step~7). Compare medians over repeated splits, as in Exercise~5.)
\item
  \textbf{Cross-fitting and coverage.} On IHDP, grade full-sample AIPW
  against cross-fit DML over 10 realizations, then 50, and reproduce the
  calibration comparison of Step~5. (With 10 realizations the gap is unstable; see the example in Step~5. At 50 realizations, the AIPW\(\to\)DML coverage gap is approximately \(0.09\) (paired \(t\approx2.8\)); this pair's bias and MSE differences are not resolved.)
\item
  \textbf{Repeat the split.} On RHC, run the fold assignment \(r=1\) vs \(r=25\) times for each of the four split-based estimators (DML, CV-TMLE, DC-AIPW, DC-TMLE) and median-aggregate (Step~5). For which estimators does the point estimate move most? (Whichever ones the first split placed farthest from their median. On these data the double-cross-fit estimates scatter more across splits; see Step~5.)
\item
  \textbf{Package reconciliation.} Run the same NHEFS estimand through \texttt{tmle} and \texttt{DoubleML} with the settings described in Step~7. Then align their learner libraries (as closely as the two ecosystems allow), their fitting schemes, and their truncation. Compare medians over 20 seeds at each stage. Which change closes most of the gap? (Record how many propensities each package's bound clips, as in Exercise~1. Single runs scatter by about one partition SD (Step~7). Comparing medians reduces dependence on a particular seed. To investigate the library and fitting scheme separately, change one setting at a time while holding the others fixed. Each change's effect is conditional on the settings held fixed, so when settings interact the sequence need not decompose the original gap uniquely.)
\end{enumerate}

\bibliographystyle{vancouvernat}
\bibliography{ref}
\end{document}